\documentclass[twocolumn,secnumarabic, 
, amssymb, aps, prl, superscriptaddress]{revtex4-1}

\usepackage{graphicx}
\usepackage{amsmath}
\begin{document}

\title{Optimal probabilistic work extraction beyond the free energy difference with a single-electron device}%

\author{Olivier Maillet}%
\email{olivier.maillet@aalto.fi}
\affiliation{QTF Centre of Excellence, Department of Applied Physics, Aalto University School of Science, P.O. Box 13500, 00076 Aalto, Finland}
\author{Paolo A. Erdman}%
\affiliation{NEST, Scuola Normale Superiore and Instituto Nanoscienze-CNR, I-56127 Pisa, Italy}
\author{Vasco Cavina}%
\affiliation{NEST, Scuola Normale Superiore and Instituto Nanoscienze-CNR, I-56127 Pisa, Italy}
\author{Bibek Bhandari}%
\affiliation{NEST, Scuola Normale Superiore and Instituto Nanoscienze-CNR, I-56127 Pisa, Italy}
\author{Elsa T. Mannila}%
\affiliation{QTF Centre of Excellence, Department of Applied Physics, Aalto University School of Science, P.O. Box 13500, 00076 Aalto, Finland}
\author{Joonas T. Peltonen}%
\affiliation{QTF Centre of Excellence, Department of Applied Physics, Aalto University School of Science, P.O. Box 13500, 00076 Aalto, Finland}
\author{Andrea Mari}%
\affiliation{NEST, Scuola Normale Superiore and Instituto Nanoscienze-CNR, I-56127 Pisa, Italy}
\author{Fabio Taddei}%
\affiliation{NEST, Scuola Normale Superiore and Instituto Nanoscienze-CNR, I-56127 Pisa, Italy}
\author{Christopher Jarzynski}%
\affiliation{University of Maryland, College Park, Maryland, 20742, USA}
\author{Vittorio Giovannetti}%
\affiliation{NEST, Scuola Normale Superiore and Instituto Nanoscienze-CNR, I-56127 Pisa, Italy}
\author{Jukka P. Pekola}%
\affiliation{QTF Centre of Excellence, Department of Applied Physics, Aalto University School of Science, P.O. Box 13500, 00076 Aalto, Finland}
\date{February 25, 2018}%
\begin{abstract}
	We experimentally realize protocols that allow to extract work beyond the free energy difference from a single electron transistor at the single thermodynamic trajectory level. With two carefully designed out-of-equilibrium driving cycles featuring kicks of the control parameter, we demonstrate work extraction up to large fractions of $k_BT$ or with probabilities substantially greater than 1/2, despite zero free energy difference over the cycle. Our results are explained in the framework of nonequilibrium fluctuation relations. We thus show that irreversibility can be used as a resource for optimal work extraction even in the absence of feedback from an external operator.
\end{abstract}
\maketitle

The ongoing miniaturization of physical systems, together with advances in techniques for the conception and manipulation of small biological objects, has made the investigation of devices with few degrees of freedom possible. In such systems fluctuations of physical quantities become comparable with or larger than their mean values. This property, in particular, has led to the theoretical \cite{jarzynski_nonequilibrium_1997,crooks_entropy_1999} and experimental \cite{wang_experimental_2002,berut_experimental_2012,collin_verification_2005} development of stochastic thermodynamics \cite{seifert_stochastic_2012}, which considers single realizations of work and heat relative to a given transformation rather than averaged quantities over an ensemble of realizations, as for the case of macroscopic systems. While the first law of thermodynamics (energy conservation) remains untouched, the second law (entropy increase over time) does not apply at the level of a single realization because of the stochastic nature of heat and work. Experimental platforms for stochastic thermodynamics include colloids \cite{berut_experimental_2012,roldan_universal_2014}, single electron boxes \cite{saira_test_2012}, electronic double dots which allow entropy production measurements \cite{kung_irreversibility_2012,singh_records_2017} and recently experiments attained the quantum regime \cite{campisi_colloquium:_2011} with e.g. NMR setups \cite{batalhao_irreversibility_2015} and superconducting circuits \cite{cottet_observing_2017,naghiloo_information_2018}. In this context, work and heat must be addressed in terms of probability distributions \cite{seifert_stochastic_2012}. In particular, work fluctuations obey the equality \cite{jarzynski_nonequilibrium_1997}
\begin{equation}
\label{Jarzynski}
\left\langle e^{-W/k_BT}\right\rangle=e^{-\Delta F/k_BT}.
\end{equation}
Here $W$ is the work \textit{performed on a system} during a single realization of the process, $\Delta F$ is the free energy difference between the system's initial and final states, $k_B$ is Boltzmann's constant and $T$ the temperature of the heat bath to which the system is connected, and angular brackets denote an ensemble average over realizations. From this equality the second law of thermodynamics is recovered, $\langle W\rangle\geq\Delta F$. Additionally, Eq. (\ref{Jarzynski}) implies that for some realizations $W<\Delta F$, i.e. the extracted work ($-W$) exceeds the decrease in free energy ($-\Delta F$). Eq. (\ref{Jarzynski}) places no limits on the magnitude of such ``violations'' of the second law, nor on the net likelihood of observing these violations. Therefore it is interesting to consider how to design a process to maximize the amount of work that might be extracted during a single realization, or alternatively to maximize the net probability to extract work beyond the free energy difference.

With the exception of recent applications of one-shot methods in this context \cite{Egloff_2015,Halpern_2015}, until now optimal control for a system coupled to a single heat bath has been mostly concerned with the trade-off between minimizing either fluctuations or average work \cite{Schmiedl2007,Solon2018}. Recently, it has been shown 
with a quantum jump approach \cite{cavina_optimal_2016} that with a suitable far-from-equilibrium driving sequence, one can instead take advantage of fluctuations to force work extraction from a system by arbitrarily large value with a non-zero probability while still obeying Eq. (\ref{Jarzynski}). In particular, Ref. \cite{cavina_optimal_2016} discusses how to perform this task in the most efficient way, finding an optimal sequence that relies on two quasi-static tuning steps of the control parameter, separated by the sudden change of its energy level spacing, also referred to as a ``quench''. Such a protocol maximizes the probability of extracting work beyond a given quantity (i.e. $W\leq W^-$ where $W^-<\Delta F$ is fixed), while ensuring that we never perform work exceeding a selected threshold $W^+$.

In this Letter, using a single electron transistor (SET) \cite{ingold_charge_1992}, we experimentally demonstrate a significant probability of extracting work arbitrarily bigger than the free energy difference in a single protocol realization. We first show, in a simple symmetric configuration of the proposed protocol, that the resulting work probability distribution follows the bounds derived in Ref. \cite{cavina_optimal_2016}, thus being optimal in the sense defined above. Building on this experimental proof we arrange the protocol in such a way that the probability of extracting work just above the free energy difference is maximized, regardless of the energy cost in case of failure. We thus observe a probability \textit{significantly greater than 1/2} of extracting work above the free energy difference, up to 65 \%, with the second law requirement $\langle W\rangle\geq\Delta F$ always satisfied. Quantitative agreement is found with both the nonequilibrium fluctuation relation [Eq. (\ref{Jarzynski})] and predictions obtained from a master equation. These results are obtained without using the information on the system's state, unlike in a ``Maxwell's demon'' \cite{koski_experimental_2014,sagawa_generalized_2010} experiment.
\\
\begin{figure}
	\includegraphics[width=4.2cm]{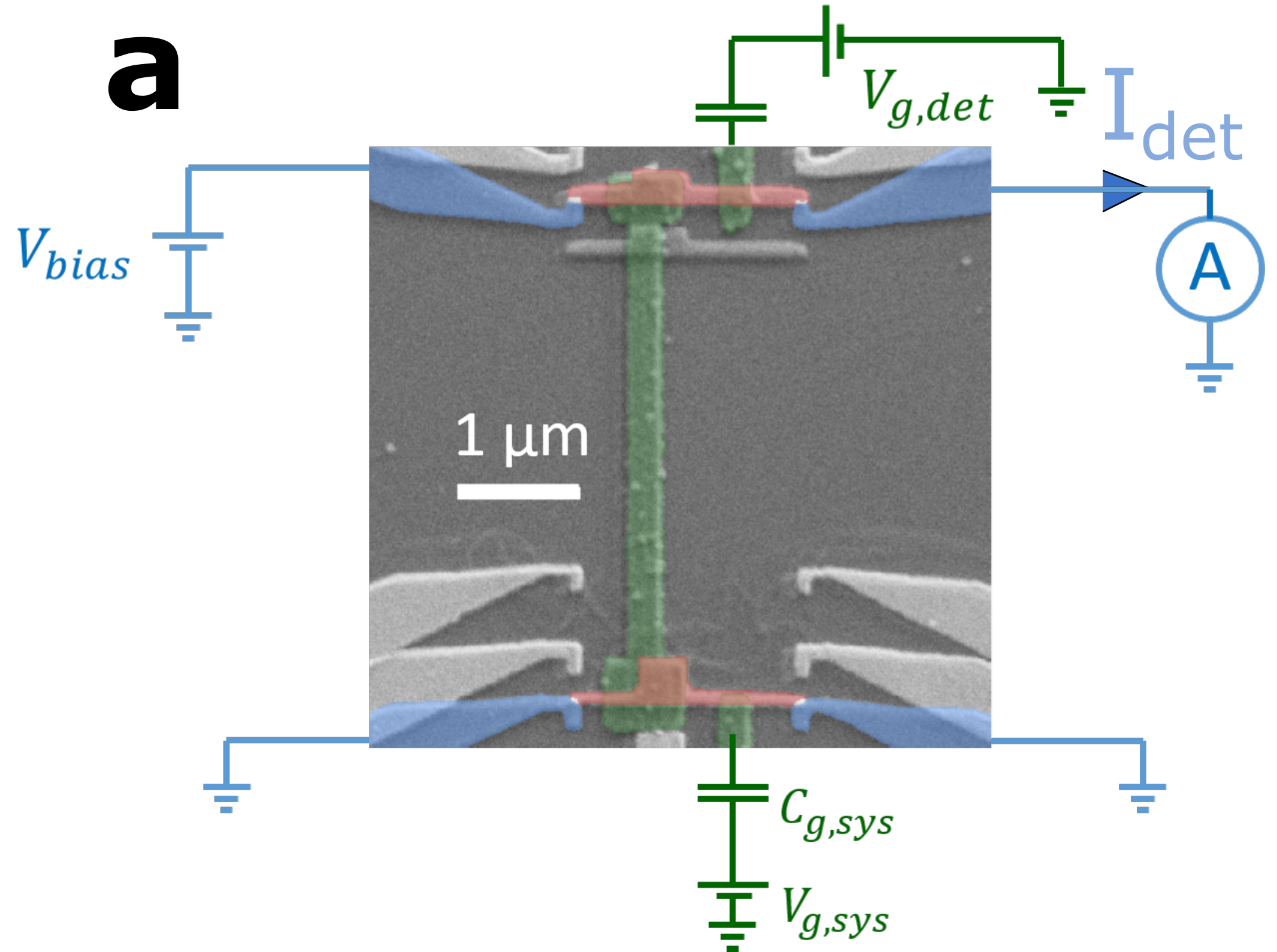}
	\includegraphics[width=4.2cm]{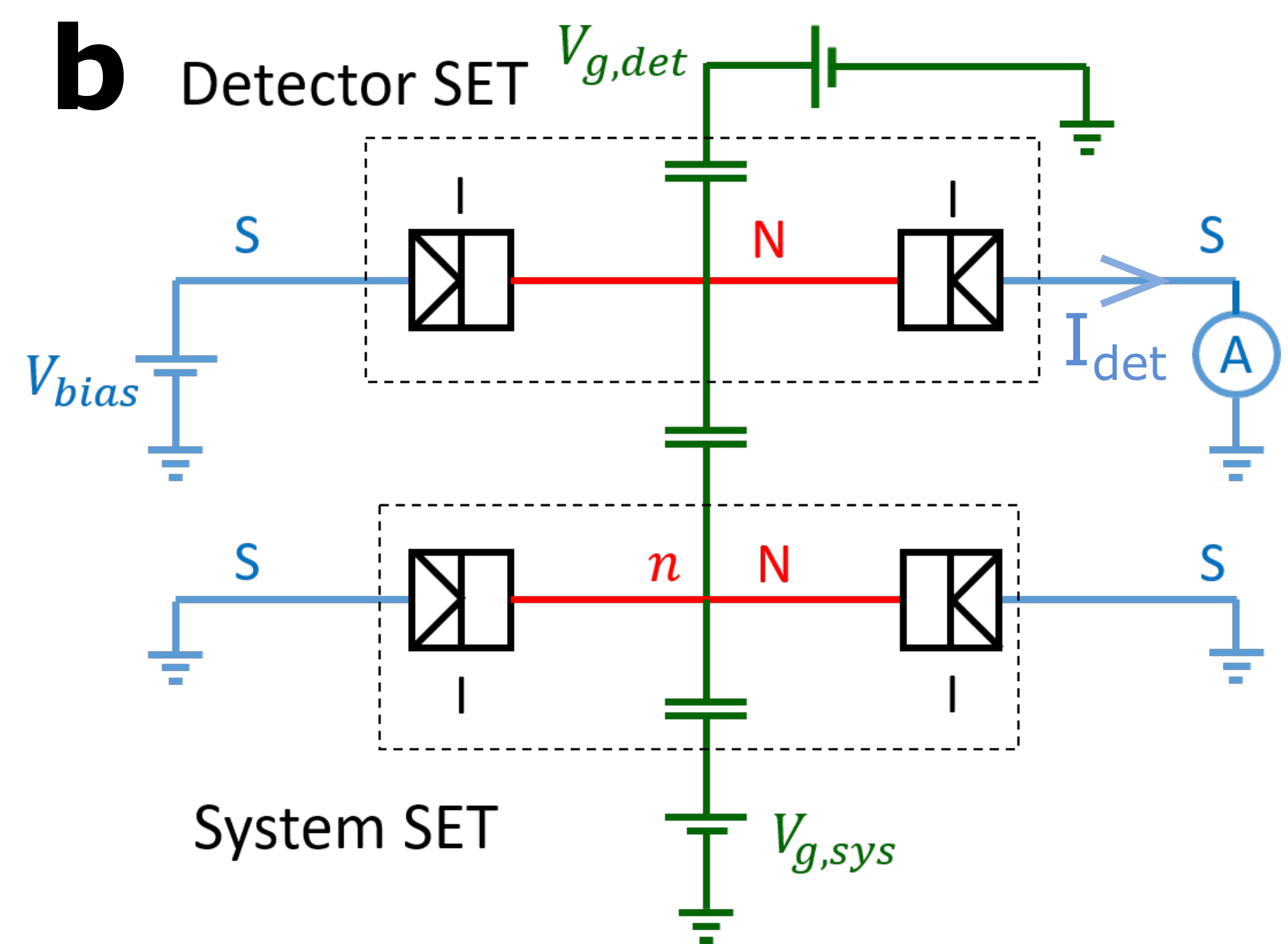}
	\includegraphics[width=8.2cm]{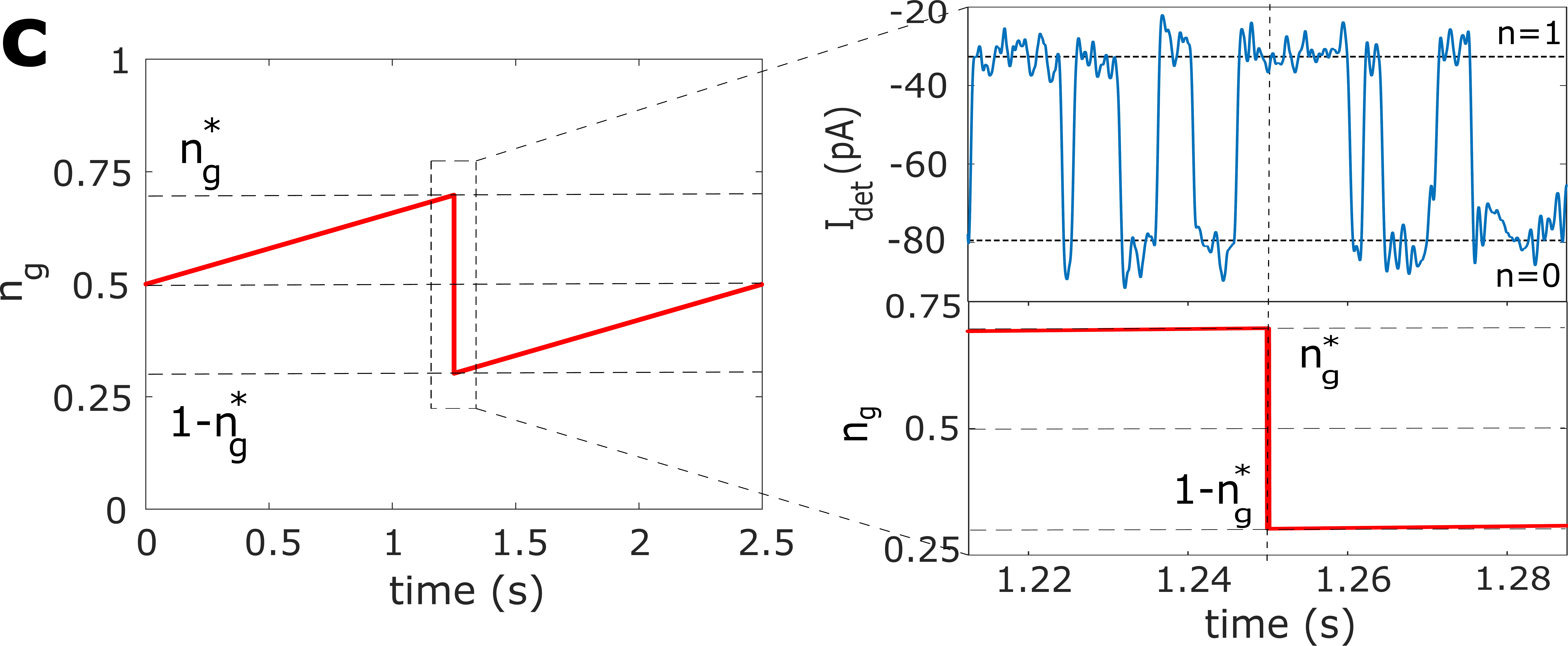}
	\caption{a) Scanning electron micrograph of the single-electron transistor (SET) capacitively coupled to a voltage biased detector SET. Leads (blue) made of superconducting aluminum are coupled through oxide (tunnel) barriers to the copper (red) island. b) Electrical circuit representation. c) Protocol used to maximize work extraction, with a zoom on the detector SET output current under system driving, around the quench event.}
	\label{fig1}
\end{figure}

The system (see Fig. \ref{fig1} a) for a micrograph and b) for a full circuit representation) is an SET fabricated through multilayer shadow evaporation \cite{fulton_observation_1987}, made of a copper island of dimensions $2000\times 200\times 25$ nm$^3$, weakly coupled through oxide tunnel barriers to superconducting aluminum leads, under zero bias. Tunnel barriers allow electron quasiparticle transport in and out of the island. Heat is carried by these electrons, and electron-electron and electron-phonon interactions take place in the island at a much faster rate than tunneling events, ensuring that a constant electronic temperature $T$ can be defined at any time \cite{giazotto_opportunities_2006}. The number $n$ of excess charges in the island is our relevant degree of freedom and the inverse tunneling rate sets the typical timescale of the system. The oxide barrier is opaque enough (the estimated tunneling resistance is $R_T\simeq 5~\mathrm{M}\Omega$ for each junction, the sum of both capacitances being $C_{\Sigma}\approx 0.7$ fF) so that its combination with superconducting reservoirs leads to low tunneling rates at zero bias, enabling measurements with a low-frequency apparatus. The electrostatic energy of the island can be tuned by an external gate voltage $V_{g,sys}$ through a gate electrode, which is patterned under the island and separated from it by a 50 nm oxide layer, forming a capacitance $C_{g,sys}=0.08$ fF $\ll C_{\Sigma}$. In this configuration the Hamiltonian of the system takes a simple form \cite{saira_test_2012},
\begin{equation}
\label{Hamiltonian}
H(n,n_g)=E_C(n-n_g)^2,
\end{equation}
where $n_g=C_{g,sys}V_{g,sys}/e$ is the reduced gate voltage and $E_C\approx e^2/2C_{\Sigma}$ is the charging energy, i.e. the energy cost of adding one electron to the island due to Coulomb interaction, which sets the energy scale of the problem. The sample is cooled down to millikelvin temperatures in a dilution refrigerator: thus, the ratio $E_C/k_B=1.3$ K is high enough so that we can restrict our analysis to two states $n=0,1$ \cite{lafarge_direct_1991-1} and the tunneling resistance is high enough to consider a sequential tunneling description. The system SET is capacitively coupled via a bottom gate electrode to another SET used as an electrometer monitoring tunneling events and hence $n(t)$. The detector SET is biased with low enough voltage so that we can modulate its output current $I_{det}$ with an external gate voltage $V_{g,det}$ between zero and (typically) 100 pA. $V_{g,det}$ is chosen to maximize the slope of current modulation $|\mathrm{d}I_{det}/\mathrm{d}V_{g,det}|$. This allows maximum sensitivity to charge variation on the system island: due to the coupling gate electrode [green vertical element in Fig. \ref{fig1}a)], electrons tunneling in or out of the system island at random times change the effective gate voltage seen by the detector SET, hence modulating its output current, which takes two values corresponding to the two charge states of the system. At charge degeneracy $n_g=1/2$, where the states $n=0$ and $n=1$ are equiprobable (no charging energy cost), these tunneling events occur at a rate $\Gamma_d=230~$Hz. This is slow enough for the detector \cite{naaman_poisson_2006}, which has a bandwidth $\sim\,1$ kHz limited by the low-pass filtering of a current amplifier. The two charge states occupation probabilities satisfy the detailed balance relation with an effective electron temperature $T=670~$mK \cite{maillet_supplementary_nodate}. From the Hamiltonian (\ref{Hamiltonian}) we know the net heat transfer $\Delta E\equiv\Delta E_{0\rightarrow 1}= H(1,n_g)-H(0,n_g)$ for an electron tunneling onto the island,
\begin{equation}
\label{heat}
\Delta E_{0\rightarrow 1}(n_g)=E_C(1-2n_g),
\end{equation}
while the opposite heat transfer for an electron leaving the island is $\Delta E_{1\rightarrow 0}(n_g)=-\Delta E_{0\rightarrow 1}(n_g)$. By monitoring tunneling events during a driving cycle, and recording the corresponding jump times $\{t_k\}$ and gate voltage values $\{n_g(t_k)\}$, we experimentally determine the total heat absorbed by the system over the thermodynamic cycle: $Q=\sum_k\Delta E[n_g(t_k)]\Delta n_k$, where $\Delta n_k=\pm 1$ depending on whether the electron jumps in/out of the island. The initial and final values of $n_g$ are both set to 1/2 so that we operate on a closed thermodynamic cycle. This way the net energy change and the free energy difference $\Delta F$ over the entire cycle are both zero, and energy conservation ensures that $W=-Q$. Thus we can directly infer the experimental value of the work at the end of the cycle based on the record of the transitions over the full cycle, see Fig. \ref{fig1}c). 

We first realize the driving sequence $n_g(t)$ depicted in Fig. \ref{fig1} c), referred to as protocol, over a time $t_f$. For a given choice of $W^-$ and $W^+$ satisfying $W^- < \Delta F < W^+$, the protocol \cite{cavina_optimal_2016} is designed to maximize the probability to observe a work value $W\leq W^-$ (successful event), while ensuring that we never observe $W^{+}\geq\Delta F$ (failure events). For the sake of simplicity we consider the symmetric case, i.e. $W^-=-W^+$. First we prepare the system at charge degeneracy, i.e. $n_g(0)=1/2$, at thermal equilibrium. Then we drive the system with a quasi-static ramp over a time $t_1\gg\Gamma_d^{-1}$ up to a value $n_g^*\equiv n_g(t_1)=1/2+\Delta n_g$, with $0<\Delta n_g<1/2$. Next, a rapid swap of the energy splitting is operated by suddenly driving the system to a value $1-n_g^*$. This ``quench'' must be realized over a time $\Delta t_q\ll \Gamma_d^{-1}$ so that no tunneling occurs in this time interval. Finally, we return the system to charge degeneracy through a quasi-static ramp, over a time $t_1$, such that $2t_1+\Delta t_q=t_f$ and $n_g(t_f)=1/2$. The total work output at the end of one cycle, obtained theoretically in the ideal quasi-static limit, writes \cite{maillet_supplementary_nodate}
\begin{equation}
\label{stochastic_work}
W(\overline{n})=(1-2\overline{n})\Delta E(n_g^*),
\end{equation}
where $\overline{n}\equiv n(t_1)$ is the charge state at the quench onset, and $\Delta E(n_g^*)<0$. Therefore $W$ is a stochastic variable taking two values $W^{\mp}=\pm\Delta E(n_g^*)$. Its distribution $P(W)=p^*\delta(W-W^+)+(1-p^*)\delta(W-W^-)$ with $1/2<p^*<1$ \cite{cavina_optimal_2016} is solely dictated by the equilibrium occupation probabilities of the two charge states before the quench, which obey the Gibbs ensemble: the ground state (one extra electron on the island) has a probability $p^*=(1+e^{\Delta E(n_g^*)/k_BT})^{-1}$, while the excited state (zero extra electron) has a probability $1-p^*=(1+e^{-\Delta E(n_g^*)/k_BT})^{-1}$. The outcome is simple to interpret physically: as the two ramps are quasi-static, the amount of work performed during those segments can be considered merely in terms of the equilibrium occupation probabilities at each instant, and is here equal to zero because of the protocol's symmetry. On the other hand, the work performed during the quench does depend on the charge state at the quench onset: if the system is in the ground state $\overline{n}=1$, the quench turns it into an energetically unfavorable state (since $\Delta E(1-n_g^*)>0$), and thus positive work has to be provided by the gate voltage source during the quench. If instead the system is in the excited state before the quench, the latter turns it into the ground state: thus energy is released by the system as work, since there is no heat exchange during the quench. Thus, counter-intuitively, the quench allows to realize $W<\Delta F=0$ by a possibly large amount by deliberately introducing irreversibility.
\begin{figure}
	\includegraphics[width=4.25cm]{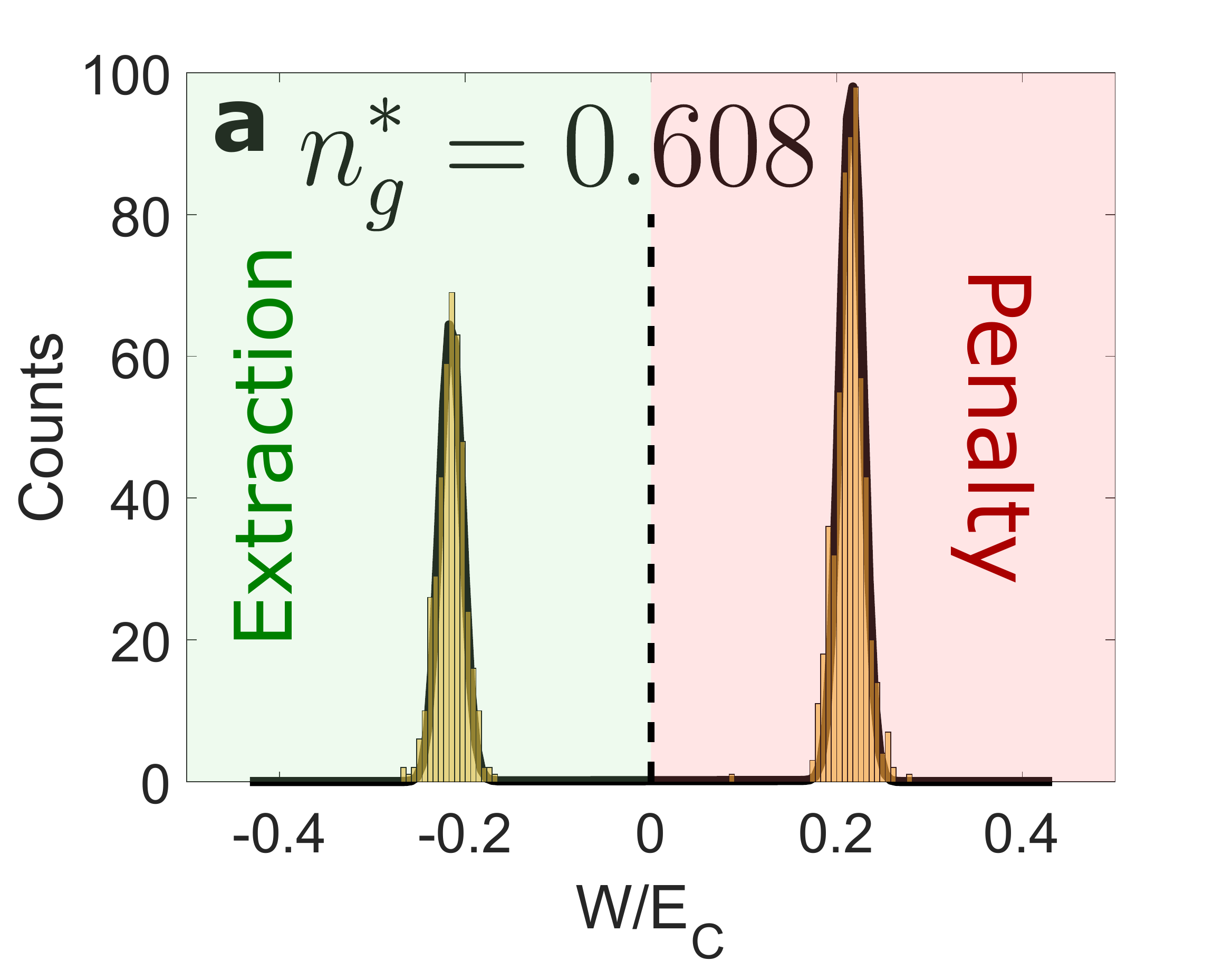}
	\includegraphics[width=4.25cm]{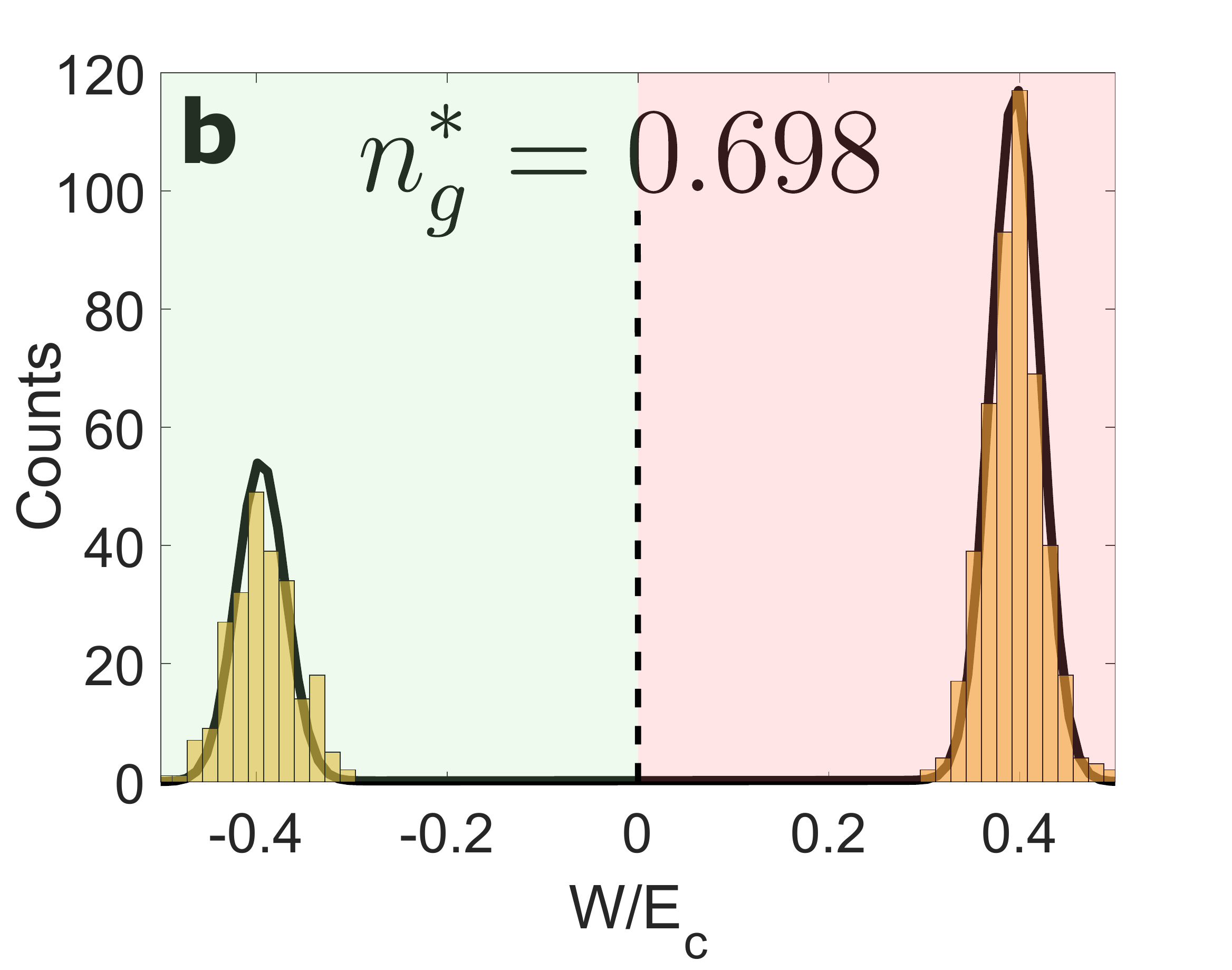}
	\includegraphics[width=4.25cm]{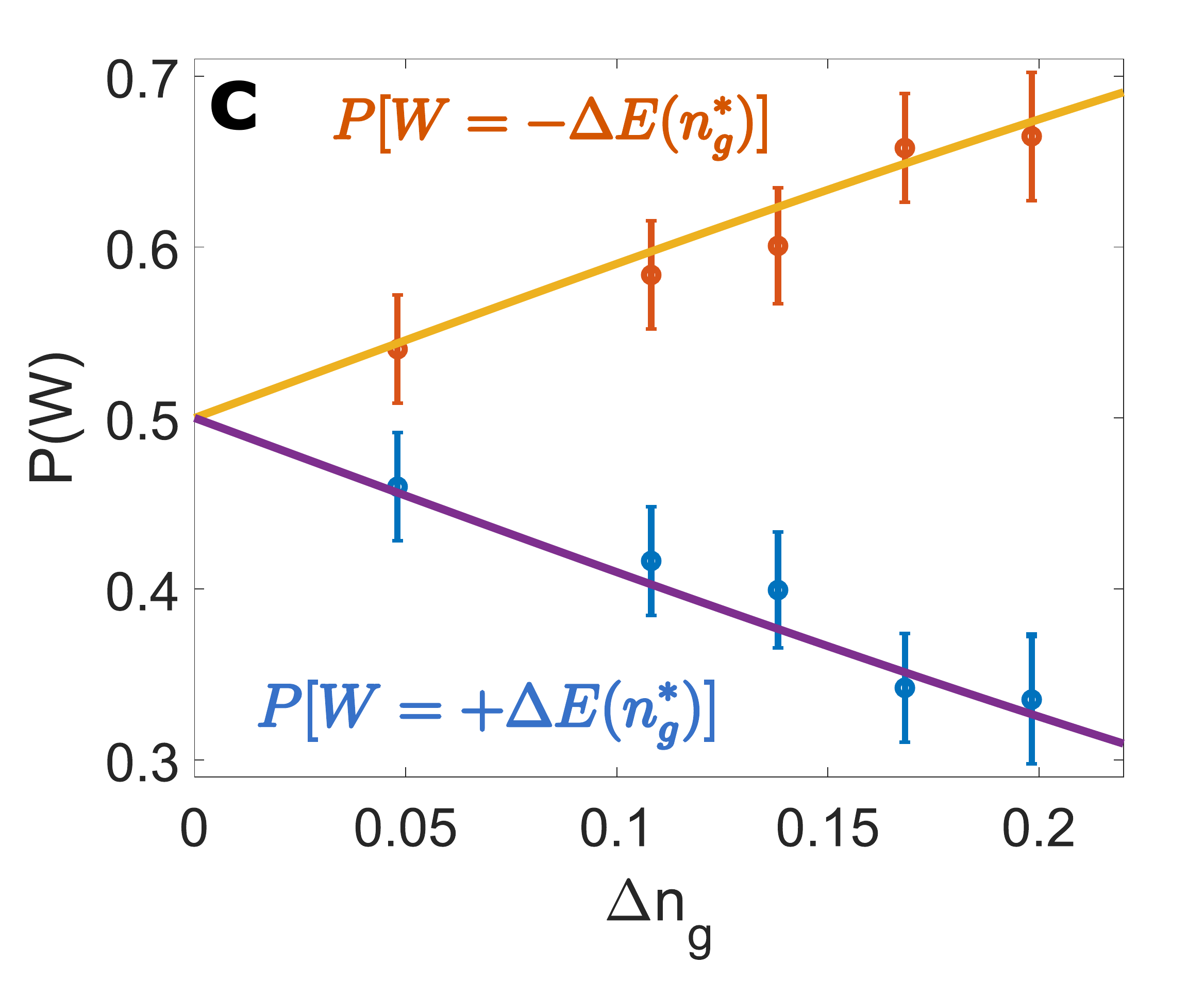}
	\includegraphics[width=4.25cm]{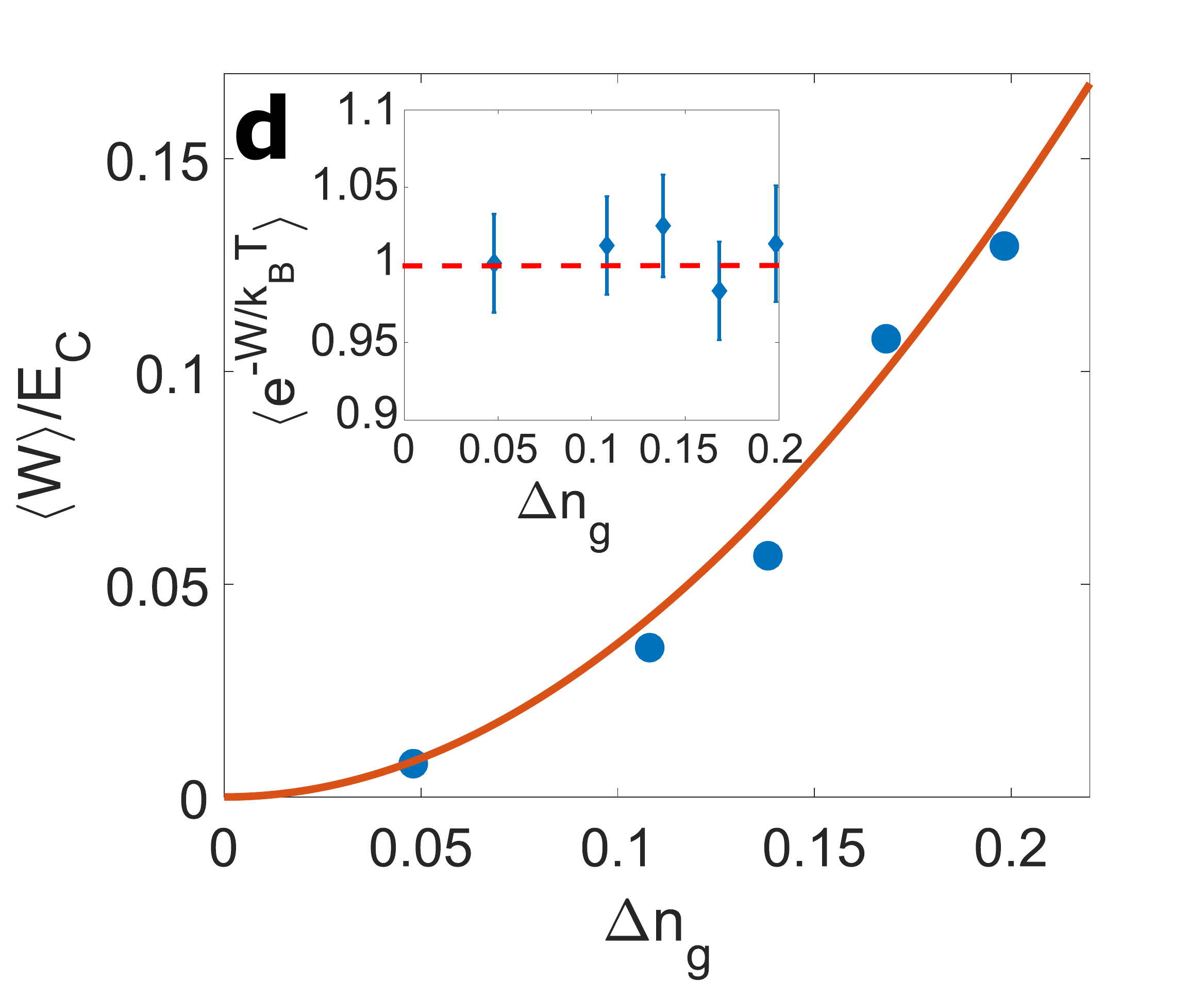}
	\includegraphics[width=4.25cm]{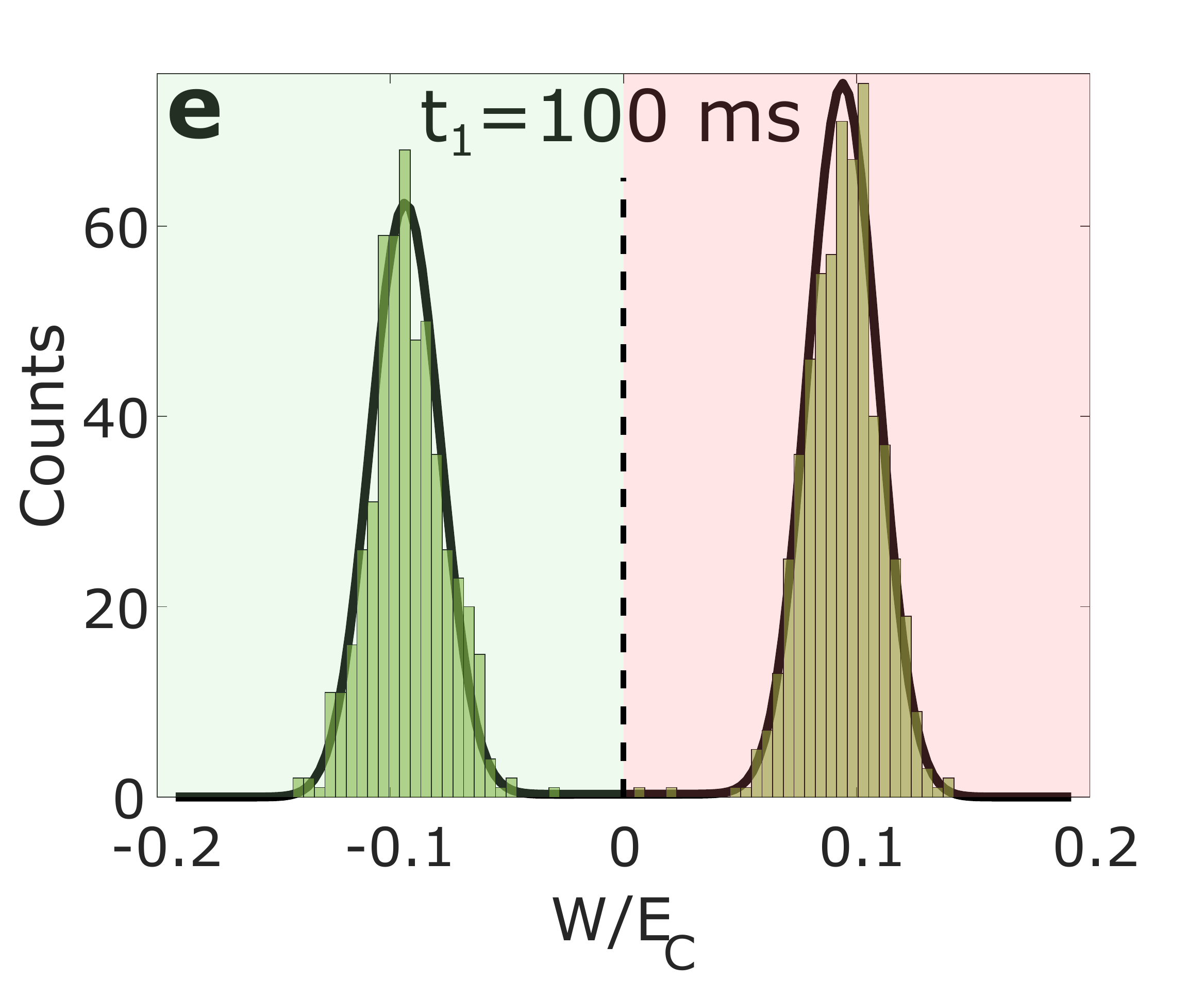}
	\includegraphics[width=4.25cm]{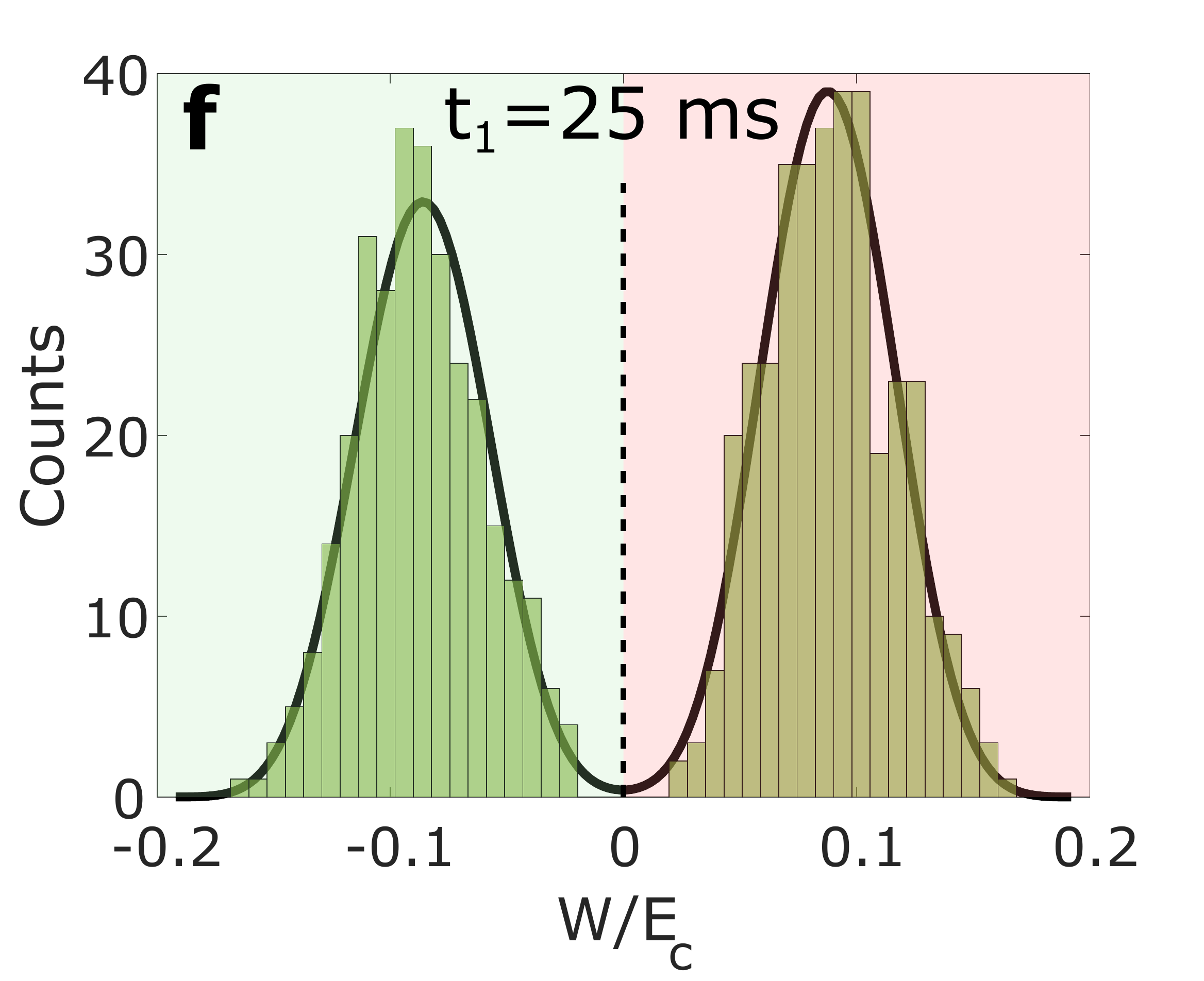}
	\caption{a) and b): work histograms obtained for a) $n_g^*=0.608$  and b) $n_g^*=0.698$, with the same ramp time $t_1=1.25~$s. c) Probability for $W=-\Delta E(n_g^*)$ (orange dots, mind the sign) and $W=\Delta E(n_g^*)$ (blue dots) events as a function of the quench amplitude. Solid lines are fits of Fermi functions (see text) with $E_C=110~\mu$eV and $T=670~$mK. Error bars are calculated from the number of protocol realizations. d) Work performed on the system averaged over all outcomes as a function of the quench peak amplitude $\Delta n_g=n_g^*-1/2$. The solid line is obtained from Eq. (\ref{avg_work}). Inset: verification of Eq. (\ref{Jarzynski}) for all values of $n_g$. e),f): work histograms obtained for the same quench amplitude $\Delta n_g=0.048$, but with ramp times $t_1=0.1$ s for e) and $t_1=0.025$ s for f), much shorter than in a). In a),b),e),f), solid lines are obtained by numerically solving the master equation \cite{maillet_supplementary_nodate}. All work values are normalized to $E_C$.}
	\label{fig2}
\end{figure}

The protocol is repeated many times ($\sim 1000$) to experimentally map the work distribution. Because of the stray capacitance associated to the electrical set-up, line filtering limits the quench time interval to $\Delta t_q=0.3$ ms, still well below $\Gamma_d^{-1}$. Work histograms obtained for two different values of $\Delta n_g$ (quench amplitudes) with the same ramp time are shown in Fig. \ref{fig2}a) and \ref{fig2}b). We indeed observe two peaks with maxima located at $\pm\Delta E(n_g^*)$. Their imbalance increases with the quench amplitude following Gibbs statistics as seen in Fig. \ref{fig2}c). This is expected since the probability $1-p^*$ to be in the excited state decreases as $n_g^*$ gets further away from charge degeneracy. Namely, the ratio between the weights of the two peaks follows the detailed balance condition for the two energy states $\pm\Delta E(n_g^*)$: $P\left[W=\Delta E(n_g^*)\right]/P\left[W=-\Delta E(n_g^*)\right]=e^{\Delta E(n_g^*)/k_BT}.$ 
Irreversibility, introduced by the quench, 
can be quantified by computing the work $\langle W\rangle=\int P(W)W\mathrm{d}W$ performed on the system, averaged over all realizations:
\begin{equation}
\label{avg_work}
\langle W\rangle=\Delta E(n_g^*)\tanh\left[\frac{\Delta E(n_g^*)}{2k_BT}\right].
\end{equation}
Indeed, $\langle W\rangle\geq 0$, as expected from the second law of thermodynamics. In Fig. \ref{fig2}d) we see that the experimental averaged work is positive and increasing with the quench amplitude, in good agreement with Eq. (\ref{avg_work}). The inset of Fig. \ref{fig2}d) shows that our work histograms obey the nonequilibrium work relation (\ref{Jarzynski}).

Note that, in contrast to the theoretical situation \cite{cavina_optimal_2016}, the peaks have a finite width in our experiment, which owes to the fact that a realistic ramp cannot be truly quasi-static, since one would need enough tunneling events between two infinitesimally close instants so that thermal equilibrium is properly defined at each instant $t$. Thus, the degree of reversibility is determined by the slope of the ramp with respect to the typical tunneling time, i.e. by $\Gamma_d^{-1}|\mathrm{d}n_g/\mathrm{d}t|$. For higher quench amplitudes but with the same ramping time, the residual irreversibility produces broader peaks \cite{saira_test_2012}, as Fig. \ref{fig2}a) and \ref{fig2}b) clearly show. We also run the protocol with constant quench amplitude but different ramp times. In Fig. \ref{fig2}e),f) work histograms for two different ramp times unambiguously demonstrate that a shorter ramp time results in a broadened distribution, as captured through a master equation approach \cite{saira_test_2012,maillet_supplementary_nodate}. 
Indeed, we see in Fig. \ref{fig2} that the obtained histograms are very well reproduced by the theoretical expectation, which validates this approach. 
\begin{figure}
	\includegraphics[width=5cm]{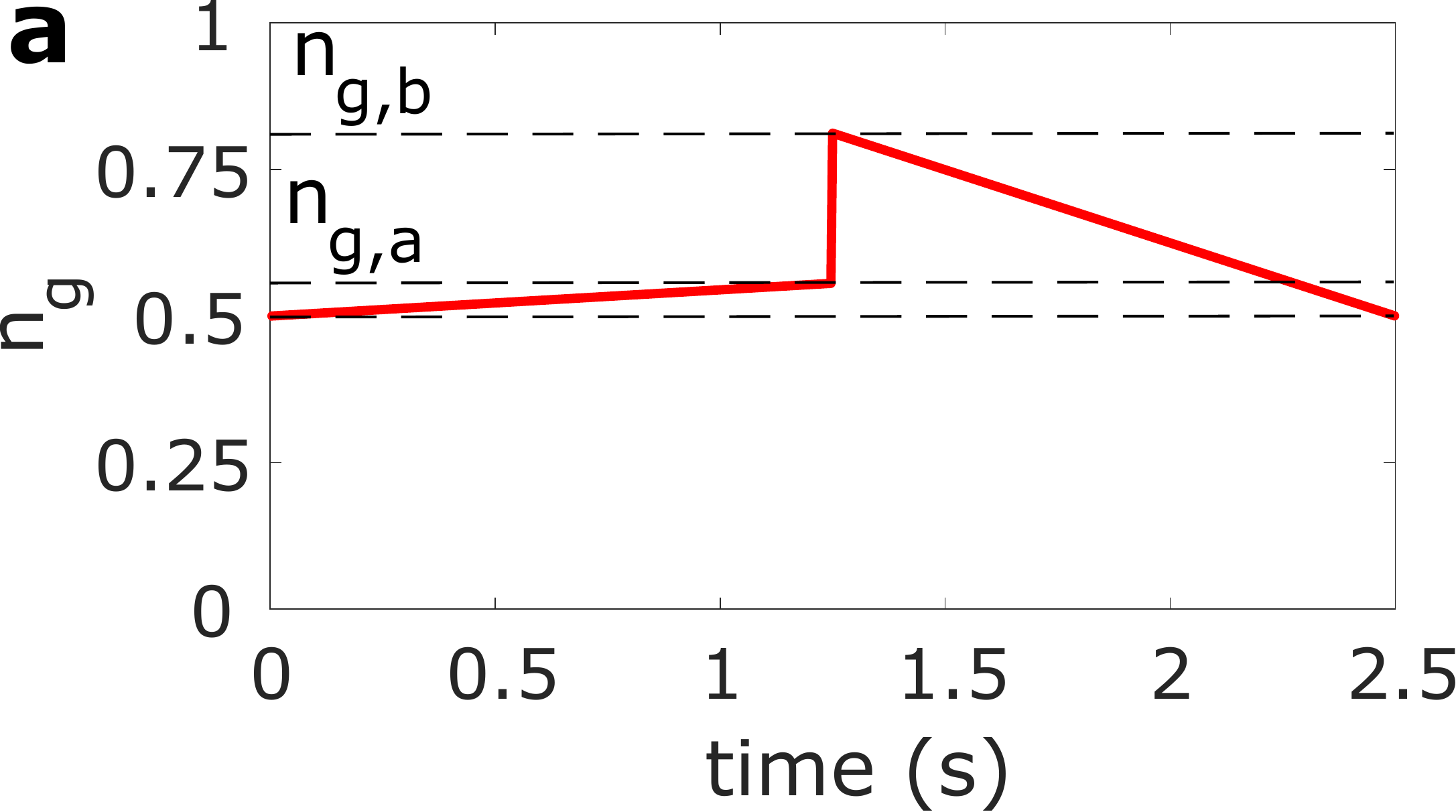}
	\includegraphics[width=4.25cm]{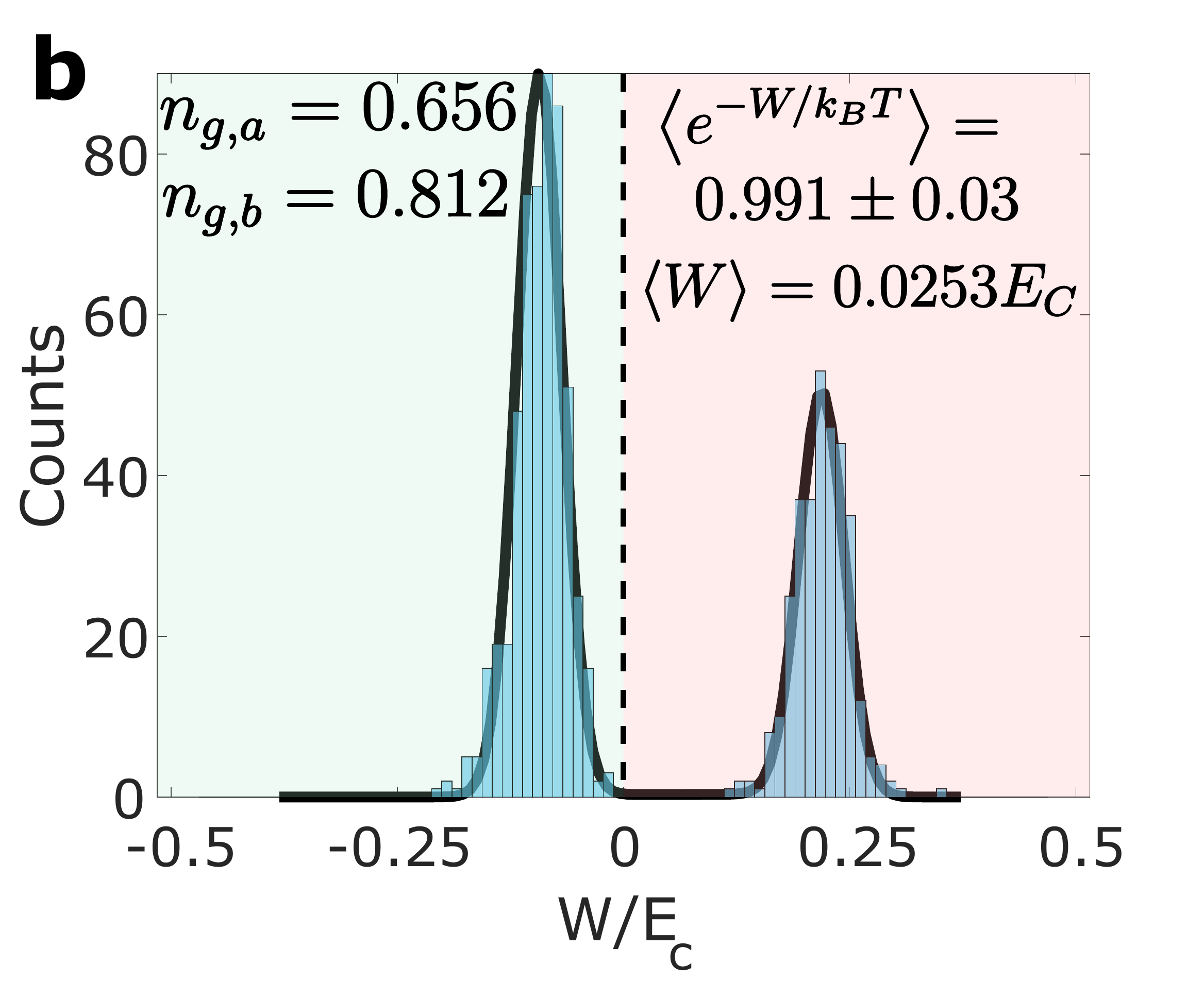}
	\includegraphics[width=4.25cm]{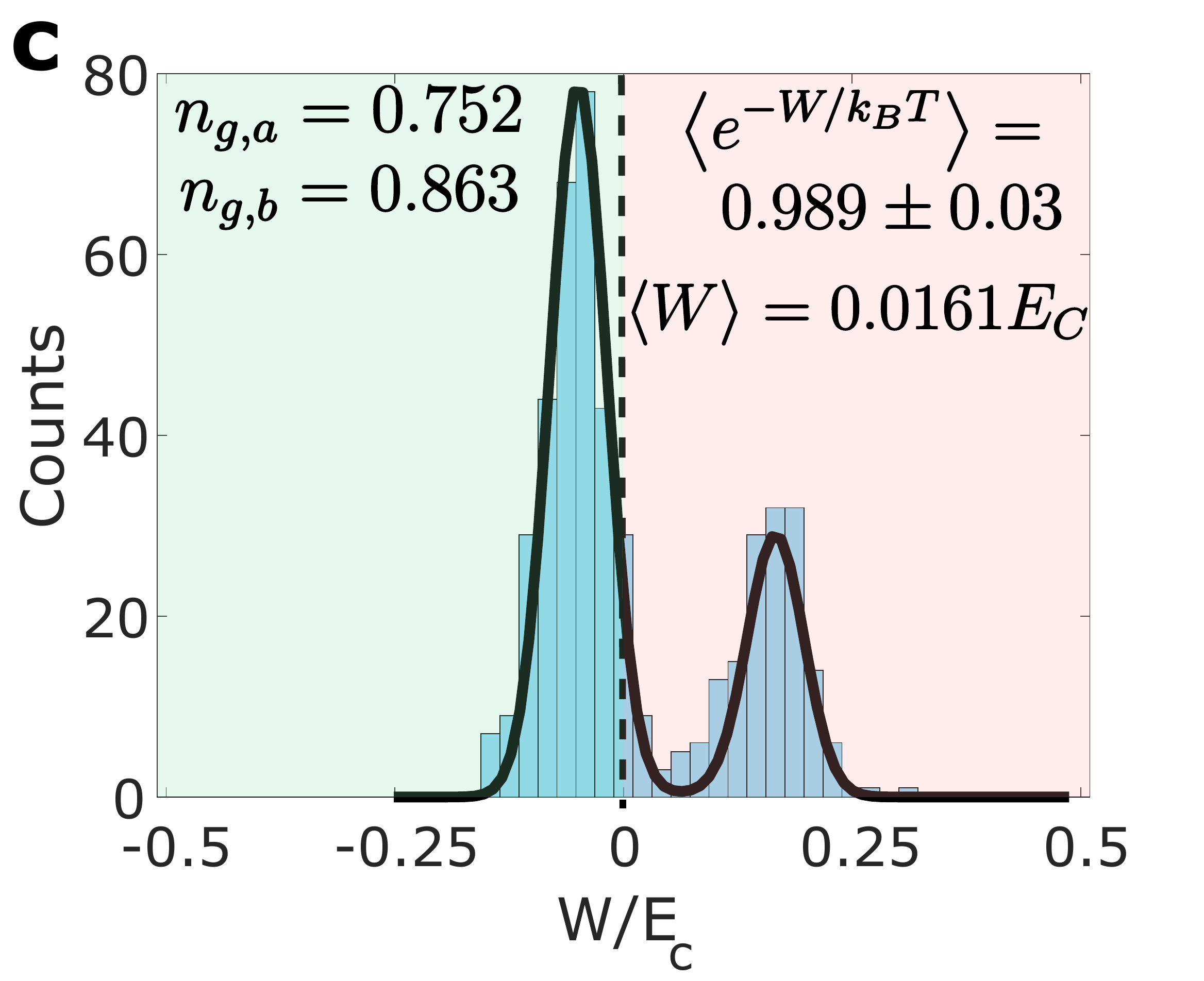}
	\caption{a) Protocol used to extract work with high probability (see text). b),c): work histograms and experimental values of work and exponentiated work averages obtained for b) $n_{g,a}=0.656,n_{g,b}=0.698$, c) $n_{g,a}=0.752,n_{g,b}=0.863$, with ramp time $t_1=1.25~$s. The vertical dashed line sets the zero free energy difference to guide the eye. Solid lines are obtained by numerically solving the master equation \cite{maillet_supplementary_nodate}.}
	\label{fig3}
\end{figure}

Next, building on this demonstration we exhibit a protocol where the goal is to maximize the probability of exceeding the second law prescription (i.e. $W<\Delta F$), without any constraint on the energy cost of the failure events. This can be achieved with the protocol depicted in Fig. \ref{fig3}a): we start at charge degeneracy, in thermal equilibrium, and ramp quasi-statically the gate voltage up to a value $n_{g,a}>1/2$. Then, in contrast with the previous protocol, we apply a quench such that the energy splitting is increased rather than reversed: over the quench time $\Delta t_q$, $n_g$ is suddenly brought to $n_{g,b}>n_{g,a}$. In the last step the system is brought back quasi-statically to charge degeneracy. With this protocol, the work performed on the system over the cycle writes \cite{maillet_supplementary_nodate},
\begin{equation}
\label{work_quench2}
W(\overline{n})=k_BT(\Delta S)_q+(p_a-\overline{n})\Delta E(n_{g,a})-(p_b-\overline{n})\Delta E(n_{g,b}),
\end{equation}
where $p_a$ ($p_b$) is the $n=1$ state equilibrium occupation probability right before (after) the quench. $(\Delta S)_q=S(p_b)-S(p_a)$ is the Shannon entropy difference between the equilibrium configurations before and after the quench, and $S(x)=-x\ln x-(1-x)\ln(1-x)$. $S$ decreases during the quench, because the splitting and occupation asymmetry become larger. For $n_{g,b}>n_{g,a}>1/2$, the sign of the work performed on the system is fully determined by the charge state at the quench onset \cite{maillet_supplementary_nodate}: $W(\overline{n}=1)<0<W(\overline{n}=0)$. Therefore, in this configuration, the probability of having $W<0$ events is determined by the ground state occupation probability $>1/2$. Indeed, if the system is in the ground state at the quench onset, the entropy decrease associated to the quench is enough to have $W<0$. In the opposite case, it is overwhelmed by the additional work required to maintain further the system in an even more unfavorable configuration. Note that this is not in contradiction with the second law of thermodynamics: from Eq. (\ref{work_quench2}), one recovers again $\langle W\rangle>0$, as confirmed experimentally together with Eq. (\ref{Jarzynski}), see Fig. \ref{fig3}b) and c). In Fig. \ref{fig3}b) an example of work histogram for such a protocol is shown. Here we indeed obtain more $W<0$ events, but such events feature small work values while $W>0$ events result in large values of work performed on the system.

In principle, there is no bound strictly below 1 to the probability of having realizations with $W<0$, since we can obtain a ground state occupation probability arbitrarily close to 1 by ramping up the gate voltage towards the Coulomb blockade regime, i.e. $n_g\rightarrow 1$ (of course, in this case the work extracted is infinitesimally small). However, achieving this is difficult in practice, because for such $n_g$, the tunneling rates from the excited to the ground state are comparable with or larger than the detector's bandwidth \cite{naaman_poisson_2006}. In addition, for reasonable ramp times, driving up to higher $n_g$ dissipates more energy. As a consequence, the peak containing $W<0$ events, which is located close to 0, broadens until the events located at the right tail of the peak are transformed into $W>0$ events, as shown in Fig. \ref{fig3}c). For such events, the irreversibility associated with an imperfect quasi-static ramp overcomes the entropy decrease due to the quench. Despite these constraints we were able to observe a probability of $65~\%$ for achieving $W<0$, still significantly greater than 1/2 [see Fig. \ref{fig3}b)].

In conclusion, we have demonstrated that a substantial amount of work can be extracted with a non negligible probability from a two-level system \textit{coupled to a single heat bath}, using a SET driven far from equilibrium with a rapid quench. The driving cycle is designed to maximize either the work or the probability of extracting work from the system on one trajectory, by strongly amplifying work fluctuations rather than minimizing them, which represents a new paradigm for work extraction in mesoscopic engines. Our experimental results satisfy the nonequilibrium work relation and agree with a master equation approach which takes into account the irreversibility associated to finite time driving.  We stress that even though work extraction can be favored, an external intervention (e.g. a Maxwell's Demon \cite{koski_experimental_2014}) would still be required to select only the extraction events: it is thanks to this absence that the second law remains valid, as we see experimentally. 
Appealing applications are foreseen if one optimizes the device: with a larger charging energy and bandwidth, using e.g. a radiofrequency detecting SET \cite{schoelkopf_radio-frequency_1998}, it should be easier to obtain either very large work extraction or work extraction probabilities very close to 1. Moreover, the deviation from the quasi-static hypothesis leaves open the question of optimizing the protocol with respect either to the work fluctuations (i.e. the peak widths) or the average values (the peaks centers). Such a problem has received a lot of theoretical attention recently: for example, it has been shown that there is an analogy with first-order phase transitions between the protocols minimizing the two quantities \cite{Solon2018}. Finally, the absence of quantum coherence in our system leaves open the question of probabilistic work extraction in the presence of quantum fluctuations and measurements \cite{cottet_observing_2017,Elouard_2017}, which could be addressed experimentally using e.g. superconducting quantum bit circuits \cite{cottet_observing_2017}.

\begin{acknowledgements}
We thank L. B. Wang for technical help, as well as S. Ciliberto, R. Fazio, S. Singh and I. M. Khaymovich for helpful discussions. This work was performed as part of the Academy of Finland Centre of Excellence program (project 310257). This work has also been supported by the SNS-WIS joint lab “QUANTRA”, and by the CNR- CONICET cooperation programme “Energy conversion in quantum, nanoscale, hybrid devices”. We acknowledge the provision of facilities by Aalto University at OtaNano-Micronova Nanofabrication Centre. C. J. acknowledges financial support from the U.S. Army Research Office under contract number W911NF-13-1-0390. 
\end{acknowledgements}
	
\bibliography{quench_paper}
\bibliographystyle{unsrt}

\onecolumngrid
\appendix
\section{Characterization}
Transport measurements have been realized on both voltage-biased system and detector SETs \cite{ingold_charge_1992}. We estimate from large voltages the series tunnel resistance $R_T=10.3$ M$\Omega$. The fit of the entire I-V characteristic yields the BCS gap $\Delta=214~\mu$eV and charging energy $E_C=109~\mu$eV $=k_B\times 1.3$ K for the system SET. The values obtained for the detector with the same procedure are $R_{T,det}=400$ k$\Omega$, $\Delta_{det}=209~\mu$eV, $E_{C,det}=83~\mu$eV. For this experiment the system SET is left unbiased and thus behaves as a single-electron box \cite{lafarge_direct_1991} with two parallel tunnel junctions, leading to an effective tunnel resistance $R_{T,//}=2.6$ M$\Omega$ seen from the box perspective. These tunnel junctions realize a weak coupling ($R_{T,//}\gg R_K=h/e^2$) of the system to the superconducting leads, which are electrically grounded. The electron quasiparticles in the superconducting leads and in the metallic island behave as a single heat bath for the two level system defined by the two charge states n=0,1 (see section 2 for details).
\section{Detailed balance, tunneling rates and effective temperature}
The energetics of the single-electron box has been addressed e.g. in Ref. \cite{pekola_work_2012}. In its reduced form, the system Hamiltonian taking into account gate driving writes:
\begin{equation}
\label{SI_Hamiltonian}
H(n,n_g)=E_C(n-n_g)^2,
\end{equation}
where $n_g=C_gV_g/e$ is the reduced gate voltage and $n$ is the number of extra electrons in the islands with respect to the chemical potential. While tunneling into the island, an electron brings heat, as there is a change in the island's energy for a given $n_g$, $\Delta E\equiv\Delta E_{0\rightarrow 1}=H(1,n_g)-H(0,n_g)$:
\begin{equation}
\label{SI_heat_one}
\Delta E(n_g)=E_C(1-2n_g).
\end{equation}
Likewise, if an electron tunnels out, a quantity $-\Delta E$ of heat is exchanged with the bath. The sign is determined by the gate charge $n_g$. At low temperatures $k_BT<E_C$ only two charge states are likely to be occupied, which we label $n=0$ and $n=1$. The tunneling rates $\Gamma^{+(-)}\equiv\Gamma_{0\rightarrow 1\,(1\rightarrow 0)}$ and occupation probabilities $p_0,p_1$ between the two charge states at equilibrium in our system satisfy the detailed balance condition:
\begin{equation}
\label{SI_detailed_balance}
\frac{p_0}{p_1}=\frac{\Gamma^-}{\Gamma^+}=\exp\left[\frac{\Delta E(n_g)}{k_BT}\right].
\end{equation}

\begin{figure}
	\centering
	\includegraphics[width=8cm]{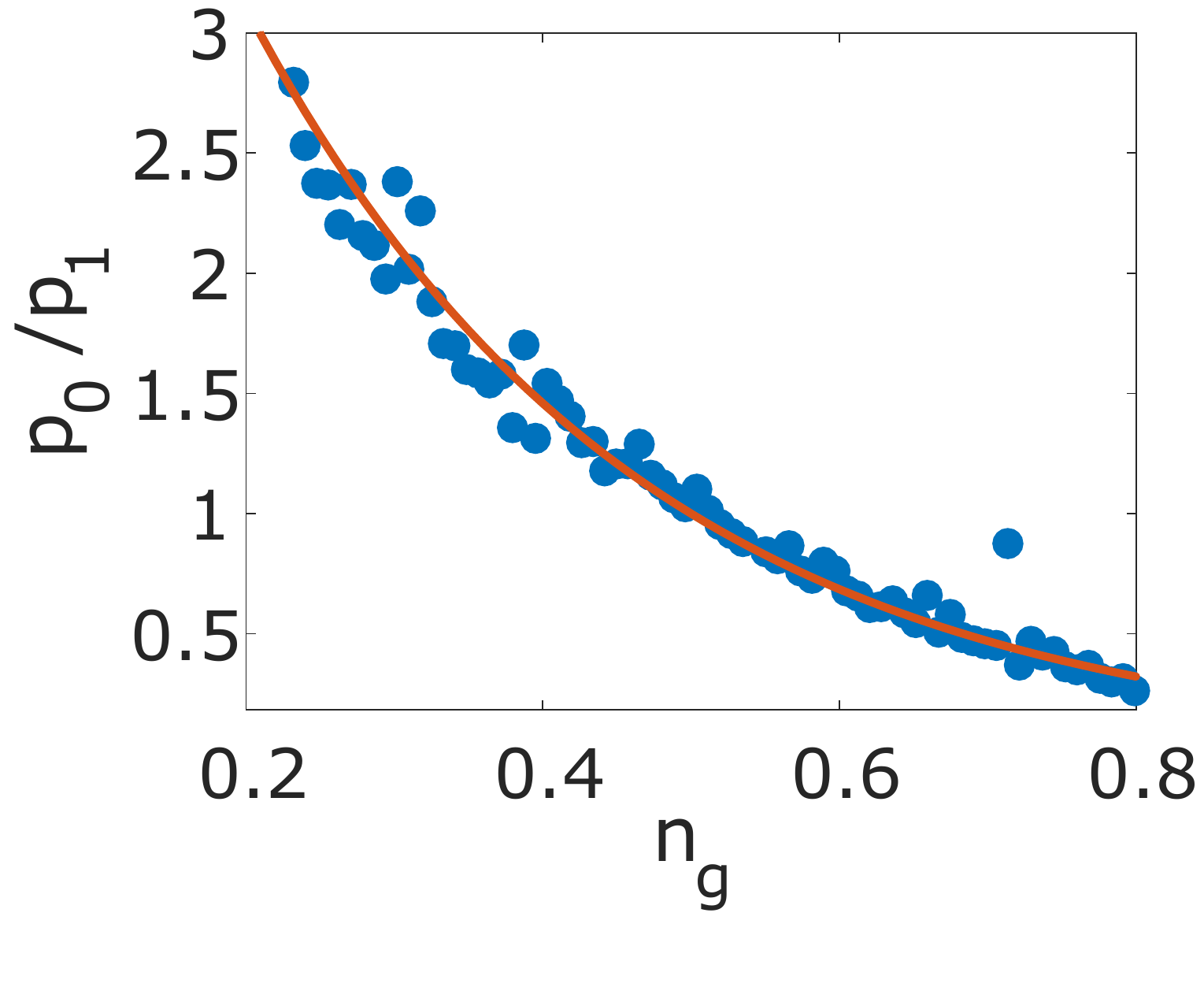}
	\caption{occupation probabilities ratio as a function of the reduced gate voltage. The red solid line is a fit using Eq. (\ref{SI_detailed_balance}) with a temperature $T_{eff}=670$ mK.}
	\label{detailed_balance}
\end{figure}

Here the rates can be determined independently using Fermi's golden rule, in the box configuration:
\begin{equation}
\label{SI_FGR}
\Gamma^{\pm}(\Delta E)=\frac{1}{e^2R_{T,//}}\int\mathrm{d}E n_S(E)f_S(E)[1-f_N(E\mp\Delta E)],
\end{equation}
where $n_S(E)=|E/\sqrt{E^2-\Delta^2}|\Theta(|E|-\Delta)$ is the BCS density of states of the superconducting leads ($\Theta$ is the Heaviside step function) and $f_{S,N}(E)=\left(1+\exp(E/k_BT_{S,N})\right)^{-1}$ is the Fermi-Dirac distribution for the superconductor and the normal island at temperature $T_S$ and $T_N$, respectively. At charge degeneracy $n_g=1/2$ we have $\Gamma^+(0)=\Gamma^-(0)=\Gamma_d$. For the purpose of the simulation later, it might be suitable to express the rates in a simpler way. As we will later confirm experimentally, the rates can be conveniently expressed, for gate charges not too far from degeneracy (i.e. $\Delta E(n_g)\ll k_B T_N$) in the form:
\begin{equation}
\label{SI_rate_simple}
\Gamma^{\pm}(\Delta E)\approx\Gamma_d\exp\left[\mp\frac{\Delta E(n_g)}{2k_BT_N}\right],
\end{equation}
which satisfies the detailed balance condition (\ref{SI_detailed_balance}), and where $\Gamma_d$ is a parameter accessible experimentally through different procedures. In the present work the bath temperature is defined through the detailed balance: we have obtained the ratio between occupancy in state 0 and state 1 as well as the rates as a function of $n_g$ (which controls the energy splitting between the two charge states), by measuring the time spent in a given state over a trace of typically 1 s $\gg \Gamma_d^{-1}$ for a given $n_g$. For $n_g<0.2$ and $n_g>0.8$ the quality of the data is altered because either $\Gamma^-$ or $\Gamma^+$ is too close to the detector's cutoff frequency of 1 kHz. Thus we restrict to $n_g\in[0.2,0.8]$, which corresponds to the range of interest for our experiment. Analyzing the equilibrium occupation probabilities (see Fig. \ref{detailed_balance}), we measure an effective temperature $T_{eff}=670\pm 30$ mK.

However, this temperature does not correspond to a physical temperature of the electrons on the normal-metal island of the system but arises from a combination of several thermal and non-thermal effects. Indeed, thermally activated tunneling rates for our devices at 670 mK would be several orders of magnitude higher than the measured rates. We stress that the results of the main text are valid as long as detailed balance is satisfied regardless of the origin of the tunneling rates. In that sense the effective temperature assigned to the system can be seen as arising from the coupling to an effective equilibrium heat bath at $T_{eff}$.
\begin{figure}[h!]
	\centering
	\includegraphics[width=8cm]{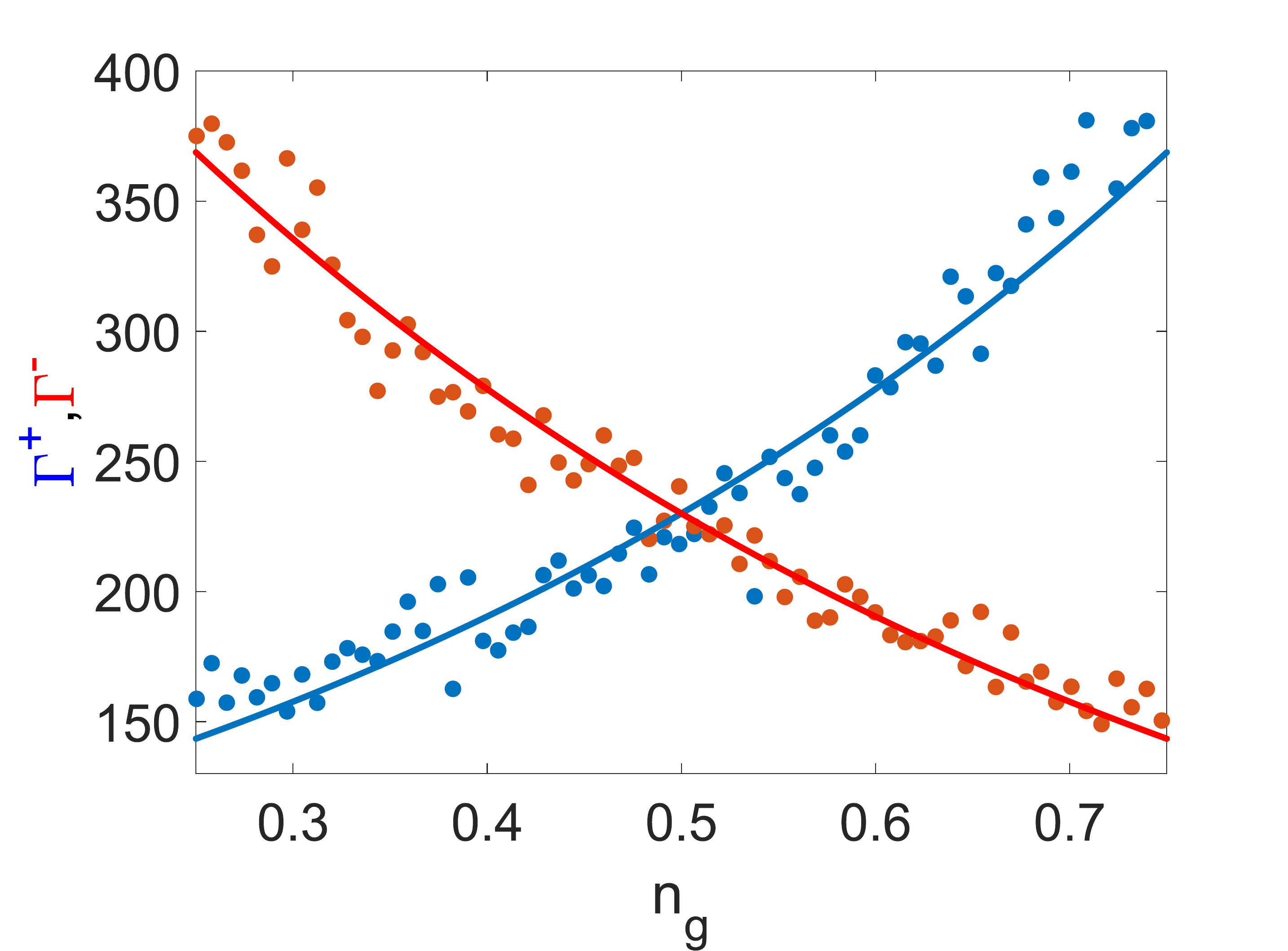}
	\caption{rates $\Gamma^{+}$ (blue) and $\Gamma^{-}$ (red) measured as a function of the reduced gate voltage $n_g$. Solid lines are fits from Eq. (\ref{SI_rate_simple}) with $T_{eff}=670$ mK and $\Gamma_d=230$ Hz.}
	\label{rates}
\end{figure}

\begin{figure}[h!]
	\centering
	\includegraphics[width=6cm]{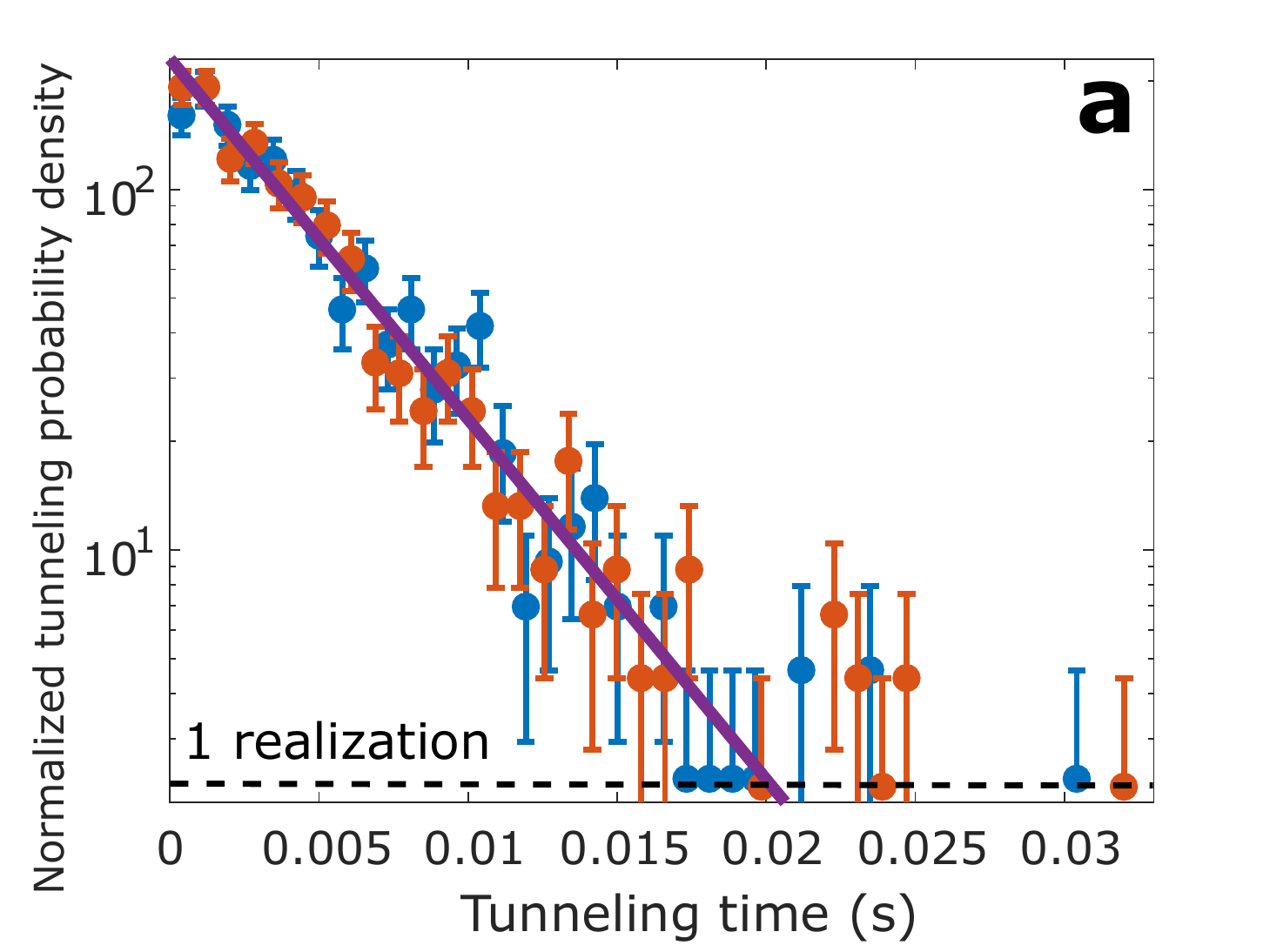}
	\includegraphics[width=6cm]{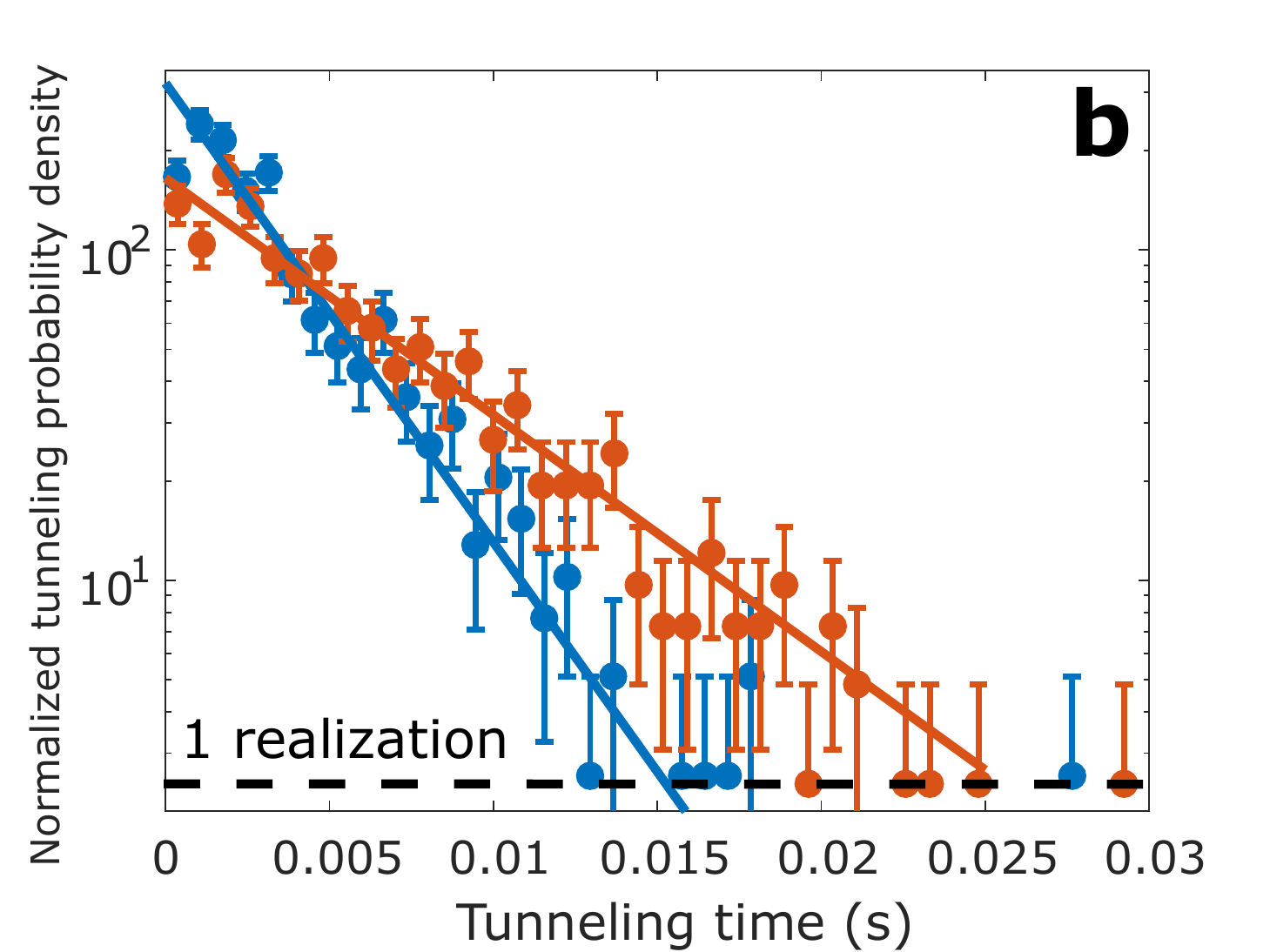}
	\caption{a) tunneling events time distribution at charge degeneracy for a given trace of duration $2$ s. The purple line is a fit following Eq. (\ref{SI_Poisson}) with $\Gamma^+=\Gamma^-=\Gamma_d=230$ Hz. b) Tunneling events time distribution at $n_g=0.7$. Solid lines yield rates $\Gamma^+=320\pm 20$ Hz (blue line) and $\Gamma^-=165\pm 20$ Hz (red line). Error bars are calculated on the basis of the number of statistical samples.}
	\label{poisson}
\end{figure}

As seen in Fig. \ref{rates}, rates are well fit in the range of interest with the expression (\ref{SI_rate_simple}) at $T_N=T_{eff}=670\pm 25$ mK, consistently with the temperature obtained from the fit of Fig. \ref{detailed_balance}. Thus, as far as our experimental range for $n_g$ is concerned, all non-thermal contributions to the rates can be recasted in $T_{eff}$ without any further effect. We obtain at $n_g=1/2$ a rate $\Gamma_d=230\pm 20$ Hz. In addition, we have checked that the tunneling events are exponentially distributed in time, as expected from a Poisson process:
\begin{equation}
\label{SI_Poisson}
\mathcal{P}_{\Gamma^{\pm}}(t)=\Gamma^{\pm}e^{-\Gamma^{\pm}t},
\end{equation}
where $t$ is the time for a tunnel event to occur. Two examples shown in Fig. \ref{poisson} for $n_g=0.5$ (charge degeneracy) and $n_g=0.7$. Note that the rate extracted from the fit at charge degeneracy (230 Hz) is consistent with the one obtained by monitoring $\Gamma^{\pm}$ as a function of $n_g$ (see Fig. \ref{rates}).

\section{Work and heat along a single trajectory}
Along a given trajectory, the total heat $Q$ exchanged is the sum over all tunneling events between the initial time $t_i$ and final time $t_f$ of the heat exchanged in each tunneling event, which rewrites in a functional, integral form:
\begin{equation}
\label{SI_heat_generic}
Q[n(t),n_g(t)]=\int_{t_i}^{t_f}\Delta E[n_g(t)]\frac{\mathrm{d}n}{\mathrm{d}t}\mathrm{d}t.
\end{equation}
Meanwhile, work in its "inclusive" definition \cite{seifert_stochastic_2012,jarzynski_comparison_2007} is associated to a variation of the Hamiltonian through the external control parameter $n_g$, referred to as "protocol" or "driving". It is a functional of the stochastic trajectory $n(t)$, and of the protocol:
\begin{equation}
\label{SI_work_generic}
W[n(t),n_g(t)]=\int_{t_i}^{t_f}\frac{\partial H}{\partial n_g}\frac{\mathrm{d}n_g}{\mathrm{d}t}\mathrm{d}t.
\end{equation}
Here $W$ is by convention the work performed by an external source (in this case, the gate voltage $V_{g,sys}$) on the system. By making explicit Eq. (\ref{SI_work_generic}) and integrating it by parts, one obtains by using Eq. (\ref{SI_heat_generic}):
\begin{equation}
\label{SI_work_explicit}
W=-Q+n(t_f)\Delta E[n_g(t_f)]-n(t_i)\Delta E[n_g(t_i)]+E_C\left[n_g^2(t_f)-n_g^2(t_i)\right].
\end{equation}
With this equality it is possible to obtain the work performed for each segment of the two protocols presented in the main text, denoted P1 and P2 by order of appearance. Note that we also recover $W=-Q$ for a cycle that starts and end at equilibrium (i.e. $n_g(t_f)=n_g(t_i)=1/2$ which is the case for P1 and P2), as a consequence of energy conservation. Besides, starting or ending at charge degeneracy simplifies Eq. (\ref{SI_work_explicit}) since $\Delta E(1/2)=0$.
\subsection{Work distribution under protocol P1}
Let us first address P1 (see main text). In the first segment, noted $a$, the system undergoes a linear ramp from $n_g(0)=1/2$ to $n_g^*\equiv n_g(t\rightarrow t_1^-)$, with $1>n_g^*>1/2$, between time $t=0$ and $t=t_1$. For a ramp time $t_1\gg\Gamma_d^{-1}$ the process is quasi-static, reversible, and in this ideal limit a few observations can be made:
\begin{itemize}
	\item The work performed during each repetition of the protocol does not fluctuate \cite{averin_statistics_2011}, so its value is equal to the thermal average of Eq. (\ref{SI_work_explicit}). This can be proven by noticing that, in the quasi-static limit, an infinite number of tunneling events occur during each infinitesimal increase $\mathrm{d}n_g$ of the gate voltage. Thus, when we compute the work performed in a single trajectory by direct integration of Eq. (\ref{SI_work_generic}), we are allowed to replace $n(t)$ (which is a fast oscillating stochastic variable on the timescale of the ramp) with its average value $p_1[n_g(t)]$, given by:
	\begin{equation}
	p_1[n_g(t)]=\frac{1}{\displaystyle\exp\left(\frac{\Delta E[n_g(t)]}{k_BT}\right)+1}.
	\end{equation}
	
	\item Since we can replace $W$ with its average value, instead of the stochastic $Q$ we can consider the average heat absorbed by the system along the quasi-static ramp. It is determined by the entropy change of the system: $Q_a=k_BT(\Delta S)_a$, where $(\Delta S)_a=S(t_1)-S(0)$ is the Shannon entropy variation between the beginning and the end of the ramp. The Shannon entropy is defined as $S=-p_1\ln p_1-(1-p_1)\ln(1-p_1)$.
\end{itemize}
Taking these aspects into account and introducing for clarity $p^*=p_1(n_g^*)$, we obtain the reversible work performed along the segment $a$:
\begin{equation}
\label{SI_workP1_a}
W_a=-k_BT(\Delta S)_a+p^*\Delta E(n_g^*)+E_C\left(n_g^{*2}-\frac{1}{4}\right).
\end{equation}
The work performed along the second quasi-static segment, noted $b$, can be derived with the same considerations. In addition, it starts after the quench at $n_g(t_1+\Delta t_q)=1-n_g^*$. Therefore we can use the fact that $\Delta E(1-n_g^*)=-\Delta E(n_g^*)$, and $p(1-n_g^*)=1-p(n_g^*)$. Thus, we have:
\begin{equation}
\label{SI_workP1_b}
W_b=-k_BT(\Delta S)_b-p^*\Delta E(n_g^*)-E_C\left(n_g^{*2}-\frac{1}{4}\right).
\end{equation}
Noticing that $(\Delta S)_b=-(\Delta S)_a$, we see that the net amount of work performed along the quasi-static segments is exactly zero, which physically is expected from the symmetry of the protocol P1. Therefore, the work produced along P1 only comes from the quench. Along the latter, of duration $\Delta t_q$, there is no tunneling event since $\Delta t_q\ll\Gamma_d^{-1}$ and thus the heat exchange is zero. Another consequence is that we have $n(t_1)=n(t_1+\Delta t_q)$. However, for the same reason, this quench drives the system far from equilibrium because it does not have time to thermalize with the heat bath along the process. Therefore the occupation number cannot be described in terms of equilibrium occupation probabilities: the work performed during the quench is truly a stochastic quantity that explicitly depends on the stochastic variable $\overline{n}\equiv n(t_1)$, that is, the charge state at the quench onset. Therefore, using Eq. (\ref{SI_work_explicit}) and equating the work along the quench to the total work performed over the cycle, we obtain the expression displayed in the main text:
\begin{equation}
\label{SI_workP1_total}
W(\overline{n})=(1-2\overline{n})\Delta E(n_g^*).
\end{equation}
Therefore $W$ can take two values $W^{\mp}=\pm\Delta E(n_g^*)$, and the ideal work probability distribution is simply:
\begin{equation}
\label{SI_workP1_distribution}
P(W)=p^*\delta(W-W^+)+(1-p^*)\delta(W-W^-),
\end{equation}
which is shown \cite{cavina_optimal_2016}, using the work fluctuation relation introduced in the main text, to be the one distribution that permits to extract maximum work above a given bound ($W\leq W^-$, where $W^-<\Delta F$ is a fixed quantity) while not performing work greater than $W^+>\Delta F$. The work performed averaged over all realizations $\langle W\rangle=\int WP(W)\mathrm{d}W$ is immediately obtained:
\begin{equation}
\label{SI_workP1_total_avg}
\langle W\rangle=(1-2p^*)\Delta E(n_g^*),
\end{equation}
which is positive for any $n_g$ between 0 and 1 (see main text).
\subsection{Work distribution under protocol P2}
We now address the second protocol P2 that allows to extract work from the system on more than half of the realizations, represented in Fig. 3 of the main text. Using the same notations as in the main text and the same arguments as for P1 for quasi-static processes, together with the convention $1/2<n_{g,a}<n_{g,b}<1$ without loss of generality (the case $0<n_{g,b}<n_{g,a}<1/2$ is symmetric and leads to the same conclusions), we can write from Eq. (\ref{SI_work_explicit}) the work performed during the two quasi-static segments:
\begin{equation}
\label{SI_workP2_a}
W_a=-k_BT(\Delta S)_a+p_a\Delta E(n_{g,a})+E_C\left(n_{g,a}^2-\frac{1}{4}\right),
\end{equation}
\begin{equation}
\label{SI_workP2_b}
W_b=-k_BT(\Delta S)_b-p_{b}\Delta E(n_{g,b})+E_C\left(\frac{1}{4}-n_{g,b}^2\right).
\end{equation}
Summing over the two quasi-static segments it is straightforward to see that $(\Delta S)_a+(\Delta S)_b=-(\Delta S)_q$, where $(\Delta S)_q$ is the difference of entropy between the initial and final states of the quench segment taken at equilibrium. Meanwhile, since there is no heat exchange during the quench, we use Eq. (\ref{SI_work_explicit}) and some simple algebra to rewrite the corresponding work performed $W_q$,
\begin{equation}
\label{SI_workP2_q} W_q(\overline{n})=E_C\left[(n_{g,b}-\overline{n})^2-(n_{g,a}-\overline{n})^2\right].
\end{equation}
Summing over all segments we obtain the expression (7) of the main text for the work performed along the protocol P2 :
\begin{equation}
\label{SI_workP2_total}
W(\overline{n})=k_BT(\Delta S)_q+(p_a-\overline{n})\Delta E(n_{g,a})-(p_b-\overline{n})\Delta E(n_{g,b}).
\end{equation}
Basic function analysis tells us that $W(\overline{n}=1)<0<W(\overline{n}=0)$, that is, work is extracted when the system is on the ground state at the quench onset, which by definition occurs with a probability greater than 1/2. Eq. (\ref{SI_workP2_total}) emphasizes that the quench brings an entropy reduction, as it moves the system further away from degeneracy, where the configurational (Shannon) entropy is maximum due to charge degeneracy, in a fast manner so there is no heat exchange contribution to the entropy variation. By close inspection of Eq. (\ref{SI_workP2_q}) we notice that work is performed on the system during the quench if the system is in the excited ($\overline{n}=0$) state before the quench, thus fighting against entropy reduction, while it is extracted in the opposite case, because the quench moves the system initially in its ground state towards a situation where the ground state occupation is even more favorable energetically.
\section{Master equation approach for a finite-time protocol}
Our experimental protocol, as in any realistic experiment, is implemented for finite times. Therefore, it intrinsically opposes to the quasi-static hypothesis made when deriving the work variable. Yet, the ideal character of this hypothesis is quantified by the ratio between the typical tunneling rate $\Gamma_d$ and the ramp slope $|\mathrm{d}n_g/\mathrm{d}t|$: the smoother the ramp, the better. In our experiment, the ramp time does not exceed 1.25 s, so as to minimize the influence of slow charge noise that introduces unwanted variations of $n_g$. Therefore a broadening of the experimental work distributions is observed because the ramp is not perfectly reversible. There is, nonetheless, a way to account for the ramp imperfection by studying the evolution of the work distribution over the protocol time. 

Let us introduce the variable $w(t)$:
\begin{equation}
\label{SI_work_reduced}
w(t)=-\int_0^tn(\tau)\frac{\mathrm{d}n_g}{\mathrm{d}\tau}\mathrm{d}\tau.
\end{equation}
Such a variable identifies with work at the end of the cycle, with a distribution $\rho(w,2t_1+\Delta t_q)=2E_CP(W)$. Since we are in the weak tunnel coupling limit $R_T\gg h/e^2,\hbar\Gamma_d\ll k_B T$, we can adopt a sequential tunneling description: we neglect higher order processes such as co-tunneling and the charge state on the island is classical. Therefore there is either one or zero extra electron in the island at a given time $t$, and we can separate the work distribution as follows:
\begin{equation}
\label{SI_work_distri_separated}
\rho(w,t)=\rho_0(w,t)+\rho_1(w,t),
\end{equation}
where $\rho_j(w,t)=\rho(w,t\cap n(t)=j)$, $j=0$ or 1. We are interested in establishing an evolution equation for $\rho_j(w,t)$ by studying small time increments $\Delta t$. Let us note $P_{i\rightarrow j}$ the probability to find the system in state $j$ at time $t+\Delta t$ knowing that it was in state $i$ at time $t$. We have:
\begin{equation}
\label{SI_rate_eq}
\begin{pmatrix} P_{0\rightarrow 0} & P_{0\rightarrow 1} \\ P_{1\rightarrow 0} & P_{1\rightarrow 1} \end{pmatrix}= \begin{pmatrix} 1-\Gamma^+(t)\Delta t & \Gamma^+(t)\Delta t \\ \Gamma^-(t)\Delta t & 1-\Gamma^-(t)\Delta t \end{pmatrix},
\end{equation}
with $\Gamma^{\pm}(t)\equiv\Gamma^{\pm}[n_g(t)]$, where we have used the expression (\ref{SI_rate_simple}) which is valid in the experimental range considered (see Fig. \ref{rates}). Let us now consider the work increment $\Delta w_{i\rightarrow j}$ associated to a transition $i\rightarrow j$ during the time interval $\Delta t$. If the system is in the state $j$ at time $t+\Delta t$, its work distribution at $t+\Delta t$ can be related to the one at time $t$ through:
\begin{equation}
\label{SI_ChapmanKolmogorov}
\rho_j(w,t+\Delta t)=\sum_{i=0,1}P_{i\rightarrow j}\rho_i(w-\Delta w_{i\rightarrow j},t).
\end{equation}
The right-hand side of Eq. (\ref{SI_ChapmanKolmogorov}) can be expanded at first order in the work increments. 
\begin{equation}
\label{SI_CK_Taylor}
\rho_j(w,t+\Delta t)\approx\sum_{i=0,1}P_{i\rightarrow j}\left[\rho_i(w,t)-\Delta w_{i\rightarrow j}\frac{\partial\rho_i}{\partial w}\right].
\end{equation}
Work increments can be made explicit using Eq. (\ref{SI_work_reduced}). Let us first notice that all the increments are first order quantities in $\Delta t$. As processes implying a change of state during $\Delta t$ are themselves first order quantities in $\Delta t$, terms including the product $P_{i\rightarrow j}\Delta w_{i\rightarrow j}$ (with $i\neq j$) are second order quantities and can be neglected when we take the $\Delta t\rightarrow 0$ limit. Therefore, we only need $\Delta w_{0\rightarrow 0}=0$ and $\Delta w_{1\rightarrow 1}=\dot{n}_g\Delta t$. These observations, together with Eq. (\ref{SI_CK_Taylor}), allow us to expand at first order this time the left-hand side of Eq. (\ref{SI_CK_Taylor}), we establish the master equation in the limit $\Delta t\rightarrow 0$:
\begin{equation}
\label{SI_mastereq_time_work}
\begin{pmatrix} \frac{\partial}{\partial t} & 0 \\ 0 & \frac{\partial}{\partial t}+\frac{\mathrm{d}n_g}{\mathrm{d}t}\frac{\partial}{\partial w} \end{pmatrix}
\begin{pmatrix} \rho_0(w,t) \\ \rho_1(w,t) \end{pmatrix}= \begin{pmatrix} -\Gamma^+(t) & \Gamma^-(t) \\ \Gamma^+(t) &-\Gamma^-(t) \end{pmatrix}\begin{pmatrix} \rho_0(w,t) \\ \rho_1(w,t) \end{pmatrix}.
\end{equation} 
From a numerical point of view, however, Eq. (\ref{SI_mastereq_time_work}) is not easily addressed. It is more convenient to work in Fourier space for the variable $w$. Let us first notice that for our finite time protocols P1 and P2,
\begin{equation}
\label{SI_upperbound_w}
|w(t)|\leq\int_0^{t}\left|\frac{\mathrm{d}n_g}{\mathrm{d}\tau}n(\tau)\right|\mathrm{d}\tau\leq\int_0^{2t_1+\Delta t_q}\left|\frac{\mathrm{d}n_g}{\mathrm{d}\tau}\right|\mathrm{d}\tau\leq\sum|\Delta n_g|\equiv w_{max},
\end{equation}
where the last sum is made over each reversible ramp segment and $\Delta n_g$ is the ramp amplitude of a given segment. Since $w$ is bounded to an interval $[-w_{max},w_{max}]$, $\rho_j$ is necessarily zero outside this interval. As such, it belongs to the category of function whose square is integrable, and thus one can expand it in a Fourier series:
\begin{equation}
\label{SI_rho_decomposed_Fourier}
\rho_j(w,t)=\frac{1}{2w_{max}}\sum_{k=-\infty}^{+\infty}\tilde{\rho}_j(k,t)e^{ikw},
\end{equation}
where $k=\pi m/w_{max}$, $m$ being an integer. The function $\tilde{\rho}_j(k,t)$ is the characteristic function, i.e. the Fourier component defined as:
\begin{equation}
\label{SI_charac_function}
\tilde{\rho}_j(k,t)=\int_{-w_{max}}^{w_{max}}\rho_j(w,t)e^{-ikw}\mathrm{d}w.
\end{equation}
Using this transformation, one can rewrite Eq. (\ref{SI_mastereq_time_work}) in a simpler form where only time derivatives are involved:
\begin{equation}
\label{SI_mastereq_time_fourier}
\frac{\partial}{\partial t}\begin{pmatrix} \tilde{\rho}_0(k,t) \\ \tilde{\rho}_1(k,t) \end{pmatrix}= \begin{pmatrix} -\Gamma^+(t) & \Gamma^-(t) \\ \Gamma^+(t) &-\left[\Gamma^-(t)+ik\frac{\mathrm{d}n_g}{\mathrm{d}t}\right] \end{pmatrix}\begin{pmatrix} \tilde{\rho}_0(k,t) \\ \tilde{\rho}_1(k,t) \end{pmatrix}.
\end{equation}
One can solve numerically Eq. (\ref{SI_mastereq_time_fourier}) for each $k$ component using a finite difference scheme: each segment is cut into small time intervals, chosen such that they are at least $10^3$ times shorter than the corresponding segment for accurate discretization. The initial distributions in $w$ space write $\rho_0(w,t=0)=(1-p_1)\delta(w)$ and $\rho_1(w,t=0)=p_1\delta(w)$. These are not "smooth" functions, simulation-wise, so we approximate them with Gaussian peaks having a small variance $\sigma^2$. Therefore, in $k$ space, the $\tilde{\rho}_j$ functions have also an initial Gaussian shape with variance $1/\sigma^2$:
\begin{equation}
\label{SI_initial_Fourier}
\left\{
\begin{array}{lll}
\tilde{\rho}_0(k,t=0) & = & (1-p_1)e^{-k^2\sigma^2/2}\\
\\
\tilde{\rho}_1(k,t=0) & = & p_1e^{-k^2\sigma^2/2}. 
\end{array}
\right.
\end{equation}
With this set of conditions, $\tilde{\rho}_j$ functions are calculated with Eq. (\ref{SI_mastereq_time_fourier}), and the final work distribution is obtained through Eq. (\ref{SI_rho_decomposed_Fourier}) taken at $t=2t_1+\Delta t_q$. Numerically, a cut-off value $k_{max}$ must be introduced. It has to be large enough to account for abrupt variations of $\rho_j(w,t)$ in $w$. Since the distribution broadens and thus becomes smoother with time, the cut-off value must be determined so that the sum (\ref{SI_rho_decomposed_Fourier}) reproduces well the initial Gaussian. Therefore, analogously to an uncertainty principle, a heuristic criterion can be formulated for a simulation to be accurate: $k_{max}\gg 1/\sigma$. However, the choice of $k_{max}$ is also upper bounded by the simulation time. We have checked by transforming an initial Gaussian distribution of width $\sigma$ according to Eq. (\ref{SI_charac_function}) and antitransforming the result according to Eq. (\ref{SI_rho_decomposed_Fourier}) by summing up to $k_{max}$ that $k_{max}\sigma\gtrsim 10$ was enough in terms of accuracy.

\begin{figure}
	\centering
	\includegraphics[width=10cm]{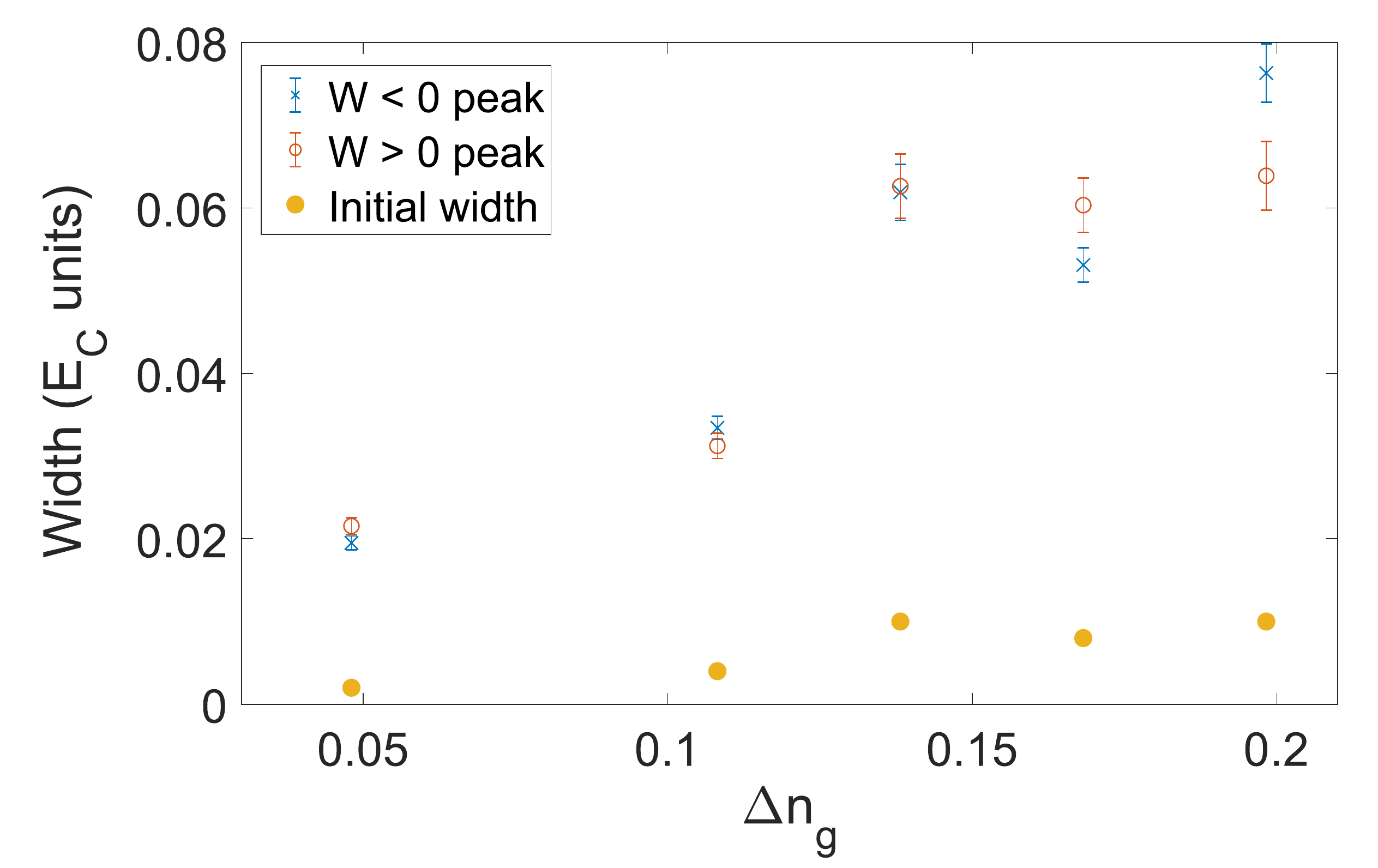}
	\caption{Measured widths of work distributions peaks for different quench amplitudes, with ramp time $t_1=1.25$ s (red and blue symbols). Error bars account for the number of realizations. Yellow dots represent the initial width $\sigma$ in the time-dependent master equation simulation.}
	\label{width}
\end{figure}

Fits are shown in the main text. We recall that in an ideal physical situation free from instrumental disturbances, the peaks width originates only from the irreversibility inherent to finite-time ramping, which yields truly physical work fluctuations. Hence, the initial width, introduced for numerical purposes and thus \textit{a priori} unphysical, should be much smaller than the final widths of the peak, and should also be independent of the ramp amplitude or time. Here, we see that the quality of the fit is influenced to some extent by the initial width chosen. However, for every fit the optimal $\sigma$ is systematically less than 16 \% of the final width. Besides, we see in Fig. \ref{width} that it roughly increases with the ramp amplitude, and we have checked that it remains essentially independent, within our measurement accuracy, from the ramp time. In fact, for shorter ramp times $t_1\ll 1$ s, the final fit is not really sensitive to the value of $\sigma$. We stress that $\sigma$ is chosen so as to optimize the fit: a reasonable agreement is still observed if $\sigma$ is taken vanishingly small.

We notice that $\sigma$ can in fact account for any broadening mechanism occuring at any time during the protocol: introducing it at $t=0$ is a mere convenience, mathematically equivalent to solving the equation with an initial Dirac and convoluting the final result with an (instrumental, non-fundamental) Gaussian noise distribution on $w$ with variance $\sigma^2$. Therefore one can speculate about a possible physical origin of $\sigma$ among a list of "instrumental" constraints:
\begin{itemize}
	\item The typical tunneling rate $\Gamma_d=230$ Hz is not infinitely smaller than the detector's bandwidth. Therefore there could be some error in the counting procedure due to "missed events": a transition, e.g. from $n=0$ to $n=1$ might be followed by the opposite transition within a time shorter than the inverse detector's bandwidth. In principle, such events  happen for two values of $n_g$ very close to each other. Therefore a bigger quench amplitude yields a bigger missed net heat transfer resulting from these two events.
	\item The expressions $\Gamma^{\pm}[n_g(t)]$ that are plugged in Eq. (\ref{SI_mastereq_time_fourier}) for the numerical solving come from the approximation (\ref{SI_rate_simple}) for numerical convenience. This becomes less valid as we ramp away from degeneracy, and in turn the obtained numerical distribution is itself an approximation which accuracy is lessened for larger $\Delta n_g$. As a result, a larger $\sigma$ might be needed for a better adjustment, even though $\sigma$ itself does not account for a peculiar physical process in this case.
	\item The signal-to-noise ratio is not necessarily optimal, in particular if the electrometer SET is not biased at the maximum slope point. Therefore spurious noise peaks might be counted as transitions by our digitizing procedure which is of trigger type.
	\item Background charge noise due to e.g. two-level-fluctuators (TLF) is responsible for slow, $1/f$-type noise on the gate charge $n_g$. Therefore it introduces an error in the sense that the computation of a tunneling event relies on a mirror driving signal sent to the computer, not on the effective, on-chip $n_g$ which is fluctuating because of these TLF. It would explain why we do not see a significant dependence for short ramp times, because then we are less sensitive to slow noise. Besides, it could explain the dependence on the quench amplitude observed for 1.25 s ramps: as one goes further away from charge degeneracy, one increases the probability of missed events such as those described above.
\end{itemize} 
\end{document}